\documentclass[aps, twocolumn, letterpaper, superscriptaddress, showpacs]{revtex4}

\usepackage{amsmath}
\usepackage{amssymb}
\usepackage{mathrsfs}
\usepackage{xspace}
\usepackage{graphicx}
\usepackage{braket}
\usepackage{color}

\usepackage{ifpdf}
\ifpdf
\fi

\newcommand{\eq}[1]{(\ref{#1})}
\newcommand{\Eq}[1]{Eq.~(\ref{#1})}
\newcommand{\Eqs}[1]{Eqs.~(\ref{#1})}

\newcommand{\Sec}[1]{Sec.~\ref{#1}}
\newcommand{\Ref}[1]{Ref.~\cite{#1}}
\newcommand{\Refs}[1]{Refs.~\cite{#1}}
\newcommand{\App}[1]{Appendix~\ref{#1}}

\newcommand{\eg}{{e.g.,\/}\xspace}
\newcommand{\ie}{{i.e.,\/}\xspace}

\newcommand{\transp}[1]{#1^{\rm T}}
\newcommand{\mc}[1]{\mathcal{#1}}
\newcommand{\mcc}[1]{\mathfrak{#1}}
\newcommand{\msf}[1]{\mathsf{#1}}
\newcommand{\mcu}[1]{\mathscr{#1}}
\renewcommand{\vec}[1]{{\boldsymbol{\rm #1}}}
\newcommand{\favr}[1]{\langle #1 \rangle}
\newcommand{\pd}{\partial}
\newcommand{\del}{\vec{\nabla}}

\begin{document}
\title{Variational principles for dissipative (sub)systems,\\ with applications to the theory of linear dispersion and geometrical optics}
\author{I.~Y. Dodin}
\affiliation{Department of Astrophysical Sciences, Princeton University, Princeton, New Jersey 08544, USA}
\affiliation{Princeton Plasma Physics Laboratory, Princeton, New Jersey 08543, USA}
\author{A.~I. Zhmoginov}
\affiliation{University of California Berkeley, Department of Physics, Berkeley, California 94720, USA}
\altaffiliation[Currently at ]{Google Inc., 1600 Amphitheatre Parkway, Mountain View, California 94043, USA.}
\author{D.~E. Ruiz}
\affiliation{Department of Astrophysical Sciences, Princeton University, Princeton, New Jersey 08544, USA}

\begin{abstract}
Applications of variational methods are typically restricted to conservative systems. Some extensions to dissipative systems have been reported too but require \textit{ad~hoc} techniques such as the artificial doubling of the dynamical variables. Here, a different approach is proposed. We show that, for a broad class of dissipative systems of practical interest, variational principles can be formulated using \textit{constant} Lagrange multipliers and Lagrangians nonlocal in time, which allow treating reversible and irreversible dynamics on the same footing. A general variational theory of linear dispersion is formulated as an example. In particular, we present a variational formulation for linear geometrical optics in a general dissipative medium, which is allowed to be nonstationary, inhomogeneous, nonisotropic, and exhibit both temporal and spatial dispersion simultaneously.
\end{abstract}

\pacs{45.20.Jj, 41.20.Jb, 42.15.-i, 52.35.-g}

% 45.20.Jj  Lagrangian and Hamiltonian mechanics
% 41.20.Jb  Electromagnetic wave propagation
% 42.15.-i  Geometrical optics
% 52.35.-g  Waves, oscillations, and instabilities in plasmas and intense beams

\maketitle

%\bibliographystyle{full}

%%%%%%%%%%%%%%%%%%%%%%%%%%%%%%%%%%%%%%%
\section{Introduction}

%--------------------------------------
\subsection{Motivation}
\label{sec:mod}

Variational methods (VM) have long been known as powerful tools in theoretical physics, especially in the context of reduced modeling. As opposed to approximating differential equations, approximating the action functional guarantees that the corresponding model too has a variational structure by definition. This implies that conservation laws are also generally preserved, so a model inherits key features of the original system notwithstanding the reduction \cite{foot:applic}. One of the areas that particularly benefit from this fact is wave theory \cite{book:whitham, book:tracy}, where even crude approximations to an action functional often yield insightful reduced models \cite{my:amc, my:wkin, my:qdirac, tex:mycovar, my:qdiel, my:qlagr, my:qdirpond, my:protation}. 

The advantages of VM become especially obvious when one deals with media such as plasmas, which can be nonstationary, inhomogeneous, anisotropic, and exhibit both temporal and spatial dispersion simultaneously \cite{my:itervar, my:sharm, my:bgk, my:trcomp, tex:myqponder, my:lens, my:autozen}. First-principle differential equations are often unrealistic to handle in this case, whereas VM allow for simple and intuitive modeling. Specifically, a wave theory can be constructed as an axiomatic field theory without even specifying the waves' nature (electromagnetic, acoustic, etc.). Defining the action functional explicitly is needed only to connect the resulting general equations with specific quantities of interest \cite{my:amc, my:wkin, my:qdirac}. For example, this provides a convenient alternative to using Maxwell's equations for electromagnetic waves \cite{my:amc}, which are generally complicated \cite{foot:comp} (as opposed to the corresponding VM) even in the geometrical-optics (GO) limit~\cite{foot:dense}. 

But which exact variational principle does one begin with? The microscopic least action principle (LAP) is typically available in the form $\delta A[q, \xi] = 0$, where $q$ describes the wave field, $\xi$ describes the medium, and $A$ is a known functional. However, practical applications in the context described above require a different variational principle, where $q$ is the \textit{only} independent variable, while the medium's degrees of freedom are ``integrated out'', \ie somehow expressed through $q$. If the medium response is adiabatic (\eg if $\xi[q]$ can be approximated with a local function), an effective LAP for $q$ can be formulated simply as $\delta A[q, \xi[q]] = 0$, as will also be discussed below. This principle has enjoyed insightful applications, \eg in the theory of generalized ponderomotive forces \cite{tex:myqponder, my:lens, my:autozen} and nonlinear plasma waves \cite{my:itervar, my:sharm, my:bgk, my:trcomp}. However, when the medium response is not adiabatic, the dynamics of $q$ becomes dissipative. (A typical example is the electric-field dynamics in Landau-damped plasma oscillations \cite{book:stix}, which we discuss in \App{app:landau}.) Then, the mentioned effective LAP becomes inapplicable, and the problem requires a more subtle approach. This issue has been recently receiving attention in plasma physics \cite{ref:benisti16} and is also important for advancing variational formulations of plasma and wave kinetics such as in \Ref{tex:myqponder}. The purpose of the present paper is to find the appropriate modification of the effective LAP that would incorporate dissipation under the above assumptions. 

%--------------------------------------
\subsection{Historical context}

Accommodating dissipative effects within the LAP has a long history (\eg cf. \Ref{ref:dekker81}) both as a formal problem of finding a variational principle for a given equation \cite{ref:bateman31, ref:jimenez76, ref:ibragimov06, ref:galley13, ref:caldirola41, ref:kanai48, ref:herrera86, ref:chandrasekar07, ref:bender07, ref:musielak08, ref:lakshmanan13, ref:riewe97, ref:rabei04, ref:frederico16, ref:bruneau02, ref:montgomery14, ref:virga15, arX:bravetti16} and also in modeling physical systems, for example, quantizing radiation in dissipative media \cite{ref:huttner92, ref:celeghini92, ref:gruner95, ref:bolivar98, ref:bechler99, ref:wonderen04, ref:bechler06, ref:suttorp07, ref:philbin10, ref:behunin11, ref:horsley11, ref:braun11, arX:churchill15}. It is hardly possible to overview the topic here, so the given references are intended as representative but not exhaustive. In any case, the subject of our paper is different from those discussed in literature, particularly, in the following aspects: 
\begin{itemize}
\item We do not address the problem of finding a LAP for a system with general dissipation. Instead, we consider only emergent dissipation in nonconservative subsystems of systems that are overall conservative. 
\item We avoid the standard approach (known under many different names \cite{ref:bateman31, ref:jimenez76, foot:dewar, ref:ibragimov06, ref:galley13}) where additional independent functions are introduced that serve as Lagrange multipliers \cite{foot:ext}. In our formulation, only time-independent Lagrange multipliers are used.
\item We avoid variable transformations that are used (as in, \eg \Refs{ref:caldirola41, ref:kanai48, ref:herrera86, ref:chandrasekar07, ref:bender07, ref:musielak08, ref:lakshmanan13}) to compensate the loss of the phase space volume caused by dissipation.
\item In contrast to theories discussed in quantum contexts, our theory is not statistical. Instead, we search for a LAP that describes deterministic dynamics on a finite time interval for a given initial state of the medium. Thus, we cannot describe wave fields using the standard Fourier representation, and we also have to specify boundary conditions (BC) in time, which are often ignored. In fact, it is the attention to these BC that helps accommodate dissipation within a LAP.
\item Instead of modeling media as collections of stationary oscillators (as in, \eg \Refs{ref:huttner92, ref:bechler06, ref:philbin10}), we allow for media that are nonstationary, inhomogeneous, anisotropic, and have both temporal and spatial dispersion simultaneously. Moreover, our approach allows treating classical and quantum degrees of freedom on the same footing.
\item Our theory is formulated within the standard calculus of variations. We avoid exotic constructs such as complex actions or fractional derivatives \cite{ref:riewe97, ref:rabei04, ref:frederico16}. Feynman integrals, which are used in related quantum theories (\eg see \Ref{ref:bechler06}), are also avoided. 
\end{itemize}

%--------------------------------------
\subsection{Main results and outline}

Our main results are as follows. We formulate a variational principle for a dissipative subsystem of an overall-conservative system by introducing constant Lagrange multipliers and Lagrangians nonlocal in time.  We call it the variational principle for projected dynamics (VPPD). Although the work is originally motivated by needs of the plasma wave theory, our results are applicable to general dynamical systems just as well. In the first part of the paper (Secs.~\ref{sec:notation}-\ref{sec:routh}), we introduce the general approach where $\xi[q]$ is an arbitrary nonlinear functional and derive an exact effective LAP for $q$. In the second part (\Sec{sec:losc}), we elaborate on applications of this approach to the important special case of linear $\xi[q]$, which corresponds to linear media. Our formulation is equally applicable to both classical and quantum media. In the third part (Secs.~\ref{sec:asym} and \ref{sec:cont}), we discuss various asymptotic approximations of the effective LAP for oscillations and waves in linear media, including the quasistatic and GO approximation. To our knowledge, this is the \textit{first variational formulation of dissipative GO} in a general linear medium, which is allowed to be nonstationary, inhomogeneous, nonisotropic, and exhibits both temporal and spatial dispersion simultaneously. Also notably, the ``spectral representation'' of the dielectric tensor $\vec{\varepsilon}_{(\omega,\vec{k})}$ that enters the GO equations is shown to be, strictly speaking, the Weyl symbol of the medium dielectric permittivity.  (Previously, it has been known that Weyl symbols emerge naturally in wave kinetics \cite{ref:mcdonald85, book:mendonca, tex:myzonal}.) This can be considered as an invariant definition of $\vec{\varepsilon}_{(\omega,\vec{k})}$, as opposed to \textit{ad~hoc} definitions used in literature that have been a source of a continuing debate \cite{ref:bornatici00, ref:bornatici03, ref:balakin15}.

Our work is also intended as a stepping stone for extending the variational theory of modulational stability in general wave ensembles \cite{tex:myqponder} to dissipative systems. Details will be described in a separate publication.

%%%%%%%%%%%%%%%%%%%%%%%%%%%%%%%%%%%%%%%
\section{Notation}
\label{sec:notation}

The following notation is used throughout the paper. The symbol $\doteq$ denotes definitions, and $\transp{(\ldots)}$ denotes transposition. For example, for a column vector $q$ [\Eq{eq:q}], one has $\transp{q} = (q_1, q_2, \ldots, q_N)$, which is a row vector; \ie in this case, $\transp{(\ldots)}$ merely lowers the index. We assume $a_n = \delta_{nm}a^m = a^n$ for any $a$, where $\delta_{nm}$ is the Kronecker delta. (Summation over repeated indices is assumed unless specified otherwise.) Hence, in principle, $a_n$ and $a^n$ could be considered equivalent, but we prefer to retain the distinction because some quantities (such as coordinates) are naturally defined with upper indexes, while others (such as momenta) are naturally defined with lower indexes. Introducing a more fundamental geometrical interpretation will not be needed for our purposes. Note also that some indexes will be used \textit{just} to introduce new symbols. This applies to all scalars (\eg $t_1$ and $t_2$) and to some vectors (\eg $q_0$, $q_T$, $\xi_0$, $\xi_0'$). We also define $a \cdot b \doteq a_n b^n$. The symbol $\pd_a$ will denote $\pd/\pd a$. The notation ``$\delta a: $'' will denote that the equation is obtained by extremizing a corresponding action functional with respect to $a$. The remaining notation is introduced within the main text. Finally, the abbreviations used in the text are summarized as follows:\\[10pt]
\begin{tabular}{@{\,} r@{\quad -- \quad} l @{\,}}
BC & boundary condition(s),\\
ELE & Euler-Lagrange equation(s), \\
GO & geometrical optics,\\
LAP & least action principle,\\
VM & variational method(s),\\
VPPD & variational principle for projected dynamics.
\end{tabular}\\

%%%%%%%%%%%%%%%%%%%%%%%%%%%%%%%%%%%%%%%
\section{Lagrangian formulation}
\label{sec:lagr}

We start by introducing the general formalism of Lagrangian mechanics. Although this formalism is commonly found in textbooks (\eg see \Ref{book:landau1}), it is essential to restate it here, so we can explicitly reference specific definitions and equations later in the text. 

%--------------------------------------
\subsection{Single system}
\label{sec:single}

Consider a Lagrangian system $\mcu{S}^{(q)}$ characterized by some real coordinate vector
\begin{gather}\label{eq:q}
q \doteq \left(
\begin{array}{c}
q^1\\
q^2\\
\vdots\\
q^N
\end{array}
\right).
\end{gather}
The dynamics of $\mcu{S}^{(q)}$ on a time interval $t \in (t_1, t_2)$ is governed by the LAP. We call this interval $T$, and we also use the symbol $T$ to denote its duration, $T \doteq t_2 - t_1$. We assume the action functional $A^{(q)}$ to have the form
\begin{gather}\label{eq:aq}
A^{(q)}[q] = \int^{t_2}_{t_1} L^{(q)}(t, q, \dot{q})\,dt,
\end{gather}
where $L^{(q)}$ is called a Lagrangian. Then, the standard formulation of the LAP is as follows:
\begin{gather}
\delta A^{(q)}[q] = 0, \\
\delta q(t_{1,2}) = 0.\label{eq:dqzero2}
\end{gather}
[Here and further, the notation $q(t_{1,2})$ means ``both $q(t_1)$ and $q(t_2)$'']. Note that \Eq{eq:dqzero2} implies BC
\begin{gather}
q(t_{1}) = {q}_0, \quad q(t_{2}) = {q}_T,\label{eq:bcq}
\end{gather}
where ${q}_0$ and ${q}_T$ are some given constants.

From \Eq{eq:aq}, one obtains
\begin{gather}\label{eq:aux1}
\delta A^{(q)} = (p \cdot \delta q)\,\big|_{t_1}^{t_2} + \int^{t_2}_{t_1}\big[\pd_q L^{(q)} - \dot{p}\big]\,\delta q\,dt,
\end{gather}
where we introduced the ``canonical momentum''
\begin{gather}
p = (p_1, p_2, \ldots, p_N)
\end{gather}
with $p_n \doteq \pd L^{(q)}(t, q, \dot{q})/\pd \dot{q}^n$, or, more compactly, $p \doteq \pd_{\dot{q}} L^{(q)}$. The first term on the right in \Eq{eq:aux1} vanishes because $q(t_{1,2})$ are fixed. Hence, the LAP leads to the following Euler-Lagrange equations (ELE):
\begin{gather}\label{eq:ele}
\delta q: \quad 0 = \pd_q L^{(q)} - \dot{p}.
\end{gather}
This is a second-order equation for the $N$-dimensional coordinate $q$, so the $2N$ BC \eq{eq:bcq} (two per each component of $q$) provide just the right number of parameters to define a solution of \Eq{eq:ele}. Alternatively, one can also approach \Eq{eq:ele} as an initial-value problem; then, the condition on $q(t = t_2)$ would need to be replaced with a condition on $\dot{q}(t = t_1)$. It is straightforward to extend this formulation to Lagrangians that depend also on higher-order derivatives, but we will not consider such extensions for the sake of brevity.

%--------------------------------------
\subsection{Coupled systems}
\label{sec:coupled}

Now suppose that $\mcu{S}^{(q)}$ is coupled to another Lagrangian system $\mcu{S}^{(\xi)}$, which we call the ``medium''. Suppose also that $\mcu{S}^{(\xi)}$ is characterized by some $M$-dimensional coordinate $\xi$. The dynamics of the resulting system $\mcu{S}$ is governed by the LAP $\delta A[q, \xi] = 0$, where $A$ is the total action. For clarity, we adopt it in the form $A = \int^{t_2}_{t_1} L\,dt$, where $L$ is the Lagrangian given by
\begin{gather}
L = 
\Lambda^{(q)} (t, q, \dot{q}) +
\Lambda^{(\xi)} (t, \xi, \dot{\xi}) 
+ \Lambda^{\rm (int)}(t, q, \dot{q}, \xi, \dot{\xi}).
\end{gather}
The assumed BC are
\begin{gather}
\delta q(t_{1,2}) = 0,\quad \delta \xi(t_{1,2}) = 0, \label{eq:dqzero}
\end{gather}
which introduce $2(M + N)$ BC similar to \Eq{eq:bcq}, namely,
\begin{gather}
q(t_{1}) = {q}_0, \quad q(t_{2}) = {q}_T, \quad \xi(t_1) = \xi_0, \quad \xi(t_2) = \xi_T.\label{eq:bcqxi}
\end{gather}
Like in \Sec{sec:single}, one then obtains
\begin{multline}\label{eq:auxA2}
\delta A[q, \xi] 
= (p \cdot \delta q)\,\big|_{t_1}^{t_2} + (\eta \cdot \delta \xi) \,\big|_{t_1}^{t_2}\\
+ \int^{t_2}_{t_1} \big[
(\pd_q L^{(q)} - \dot{p})\,\delta q + 
(\pd_\xi L^{(\xi)} - \dot{\eta})\,\delta \xi\,
\big]\,dt,
\end{multline}
where we introduced 
\begin{gather}
L^{(q)} \doteq \Lambda^{(q)} + \Lambda^{\rm (int)}, \quad 
L^{(\xi)} \doteq \Lambda^{(\xi)} + \Lambda^{\rm (int)}, \\ 
p \doteq \pd_{\dot{q}} L^{(q)}, \quad \eta \doteq \pd_{\dot{\xi}} L^{(\xi)}.\label{eq:petadef}
\end{gather}
The first two terms on the right side of \Eq{eq:auxA2} vanish due to \Eqs{eq:dqzero}, so the LAP yields the following ELE:
\begin{gather}
\delta q: \quad 0 = \pd_q L^{(q)} - \dot{p} \equiv F^{(q)},\label{eq:ele21}\\
\delta \xi: \quad 0 = \pd_\xi L^{(\xi)} - \dot{\eta} \equiv F^{(\xi)}.\label{eq:ele22}
\end{gather}
Equations \eq{eq:ele21} and \eq{eq:ele22} are second-order equations for $q(t)$ and $\xi(t)$, so the mentioned BC provide just enough parameters to define a solution.

%--------------------------------------
\subsection{Projected dynamics}
\label{sec:constrained}

Although \Eqs{eq:ele21} and \eq{eq:ele22} form a self-consistent system, it is also convenient to approach \Eq{eq:ele22} formally as if $q$ were as a prescribed function. In that case, one can, in principle, solve for $\xi$ and express it as some functional $\hat{\xi} = \xi[q]$, provided that one is given a $2M$-dimensional integration constant $C$ to determine a solution unambiguously. The resulting system is characterized by $q$ alone and is not conservative. Our goal is to derive an effective variational principle for this subsystem, or, in other words, for the dynamics of $\mcu{S}$ ``in projection'' on $\mcu{S}^{(q)}$.

First, consider the variation of $A$ evaluated on $[q, \hat{\xi}]$:
\begin{gather}
\delta A[q, \hat{\xi}] = (p \cdot \delta q)\, \big|_{t_1}^{t_2} + (\hat{\eta} \cdot \delta \hat{\xi}) \,\big|^{t_2}_{t_1} + \delta \mcu{F}^{(q)} + \delta \mcu{F}^{(\xi)},\\
\delta \mcu{F}^{(q)} \doteq \int^{t_2}_{t_1} F^{(q)}(t, q, \dot{q}, \hat{\xi}, \dot{\hat{\xi}})\,\delta q\,dt,\notag\\
\delta \mcu{F}^{(\xi)} \doteq \int^{t_2}_{t_1} F^{(\xi)}(t, q, \dot{q}, \hat{\xi}, \dot{\hat{\xi}})\,\delta \xi\,dt.\notag
\end{gather}
By definition of $\hat{\xi}$, one has $\delta \mcu{F}^{(\xi)} = 0$. We will also adopt $\delta q(t_{1,2}) = 0$, as usual. Then,
\begin{gather}
\delta A[q, \hat{\xi}] = (\hat{\eta} \cdot \delta \hat{\xi})\,\big|^{t_2}_{t_1} + \delta \mcu{F}^{(q)}.\label{eq:A10}
\end{gather}
If $C$ is a local function of $(\xi_0, \xi_T)$, then the first term on the right of \Eq{eq:A10} vanishes. In this case, by requiring that $A$ is extremized, one arrives at the correct equation \eq{eq:ele21}. However, such choice of $C$ is inconvenient. It is more practical to choose $C$ such that it characterizes the \textit{initial} state of the medium, namely,
\begin{gather}\label{eq:bcxi}
\hat{\xi}(t_1) = \xi_0, \quad \dot{\hat{\xi}}(t_1) = \xi'_0.
\end{gather}
[For example, if $q$ is electric field, and $\xi$ is linear polarization, then $\smash{\hat{\xi}} = X(t, \xi_0, \xi_0') + \smash{\int^t_{t_1} \varkappa(t, t')\,q(t')\,dt'}$, where $X$ describes free oscillations independent of $q$, and $\varkappa$ is some kernel (Secs.~\ref{sec:losc}-\ref{sec:cont}).] In this case, $\smash{\delta \hat{\xi}(t_1)}$ is zero, but $\smash{\delta \hat{\xi}(t_2)}$ is not even though $q(t_{1,2})$ are fixed. Hence,
\begin{gather}
\delta A[q, \hat{\xi}] = (\hat{\eta} \cdot \delta \hat{\xi})\big|^{t_2} + \delta \mcu{F}^{(q)},\label{eq:A1}
\end{gather}
where the first term on the right generally does not vanish. [An exception is the special case when $\smash{\hat{\xi}}$ is a local function of $q$; then, fixing $q(t_2)$ implies fixing $\smash{\hat{\xi}(t_2)}$ too.] This means that satisfying \Eq{eq:ele21} at fixed $q(t_{1,2})$ is \textit{not} sufficient to extremize $\smash{A[q, \hat{\xi}]}$. In order to ensure that the action is extremized at fixed $q(t_{1,2})$, we must constrain the value of $\smash{\hat{\xi}}(t_2)$ separately. 

The problem of extremizing $A$ under an additional constraint $\smash{\hat{\xi}}(t_2) = \text{const}$ is a standard isoperimetric problem that is solved by introducing a Lagrange multiplier \cite[Sec.~II.14]{book:lanczos}. In our case, it is an $M$-dimensional multiplier, which we will call $\lambda$. [The fact that $\lambda$ itself needs to be found increases the number of independent variables in the system, which is why $\hat{\xi}(t_2)$ can be constrained independently from $q(t_{1,2})$.] Specifically, the dynamics ``projected'' on $\mcu{S}^{(q)}$ is governed by the ``effective'' action
\begin{gather}\label{eq:aeff}
A^{\rm (eff)}[q] \doteq A[q, \hat{\xi}] + \lambda \cdot \hat{\xi}(t_2),
\end{gather}
and the corresponding variational principle, VPPD,~is
\begin{gather}\label{eq:bceff}
\delta A^{\rm (eff)}[q] = 0, \quad \delta q(t_{1,2}) = 0.
\end{gather}

The value of $\lambda$ can be readily found in the general case as follows. Under the BC \eq{eq:bceff}, one has
\begin{gather}
\delta A^{\rm (eff)}[q] = \delta \mcu{F}^{(q)} + (\eta_T + \lambda) \cdot \delta \hat{\xi}(t_2).
\end{gather}
Here, $\eta_T$ is the value of the functional $\hat{\eta}(t_2)$ corresponding to the physical trajectory for given $q_0$, $q_T$, $\xi_0$, and $\xi'_0$. The fact that the assumed fixed value of $\hat{\xi}(t_2)$ must be consistent also with the assumed $q(t_2)$ implies $\delta \mcu{F}^{(q)} = 0$. Hence, to satisfy \Eq{eq:bceff}, we adopt
\begin{gather}\label{eq:aux71}
\lambda = - \eta_T.
\end{gather}
This ensures that $\delta A^{\rm (eff)}[q] = 0$, so the variational principle \eq{eq:bceff} indeed leads to the correct equation~\eq{eq:ele21}. 

The VPPD \eq{eq:bceff}, combined with the solution for $\lambda$ given by \Eq{eq:aux71}, is the main conceptual result of this paper. In what follows, we discuss various modifications of this principle and its applications to specific types of problems of practical interest.

%%%%%%%%%%%%%%%%%%%%%%%%%%%%%%%%%%%%%%%
\section{Routhian formulation}
\label{sec:routh}

%--------------------------------------
\subsection{Basic equations}

First, consider the following alternative formulation. We start by introducing the Hamiltonian of $\mcu{S}^{\xi}$,
\begin{gather}\label{eq:hxi}
H^{(\xi)} \doteq \eta \cdot \dot{\xi}  - L^{(\xi)},
\end{gather}
as a function of $(t, q, \dot{q}, \xi, \eta)$. At fixed $(t, q, \dot{q})$, one has
\begin{align}
dH^{(\xi)}
& = d\eta \cdot \dot{\xi} + \eta \cdot d\dot{\xi} - (\pd_\xi L^{(\xi)}) \cdot d\xi - (\pd_{\dot{\xi}} L^{(\xi)}) \cdot d\dot{\xi} \notag \\
& = d\eta \cdot \dot{\xi} + \eta \cdot d\dot{\xi} - \dot{\eta}\cdot d\xi - \eta \cdot d\dot{\xi} \notag\\
& = d\eta \cdot \dot{\xi} - \dot{\eta} \cdot d\xi,
\end{align}
where we substituted \Eqs{eq:petadef} and \eq{eq:ele22}. This leads to
\begin{gather}\label{eq:he}
\dot{\xi} = \pd_\eta H^{(\xi)},\quad \dot{\eta} = - \pd_\xi H^{(\xi)},
\end{gather}
which are known as Hamilton's equations. They are equivalent to the combination of \Eqs{eq:petadef} and \eq{eq:ele22}, as one can also recheck by direct calculation using \Eq{eq:hxi}.

Consider rewriting $A$ in terms of these new variables:
\begin{gather}\label{eq:apq}
A = \int^{t_2}_{t_1} \big[\Lambda^{(q)}(t, q, \dot{q}) + \eta \cdot \dot{\xi} - H^{(\xi)}(t, q, \dot{q}, \xi, \eta)\big]\,dt,
\end{gather}
where $- \Lambda^{(q)} + H^{(\xi)}$ can be recognized as a Routhian \cite[Sec.~41]{book:landau1}. Then, 
\begin{gather}
\delta A = p \cdot \delta q\,\big|_{t_1}^{t_2} + \eta \cdot \delta \xi\,\big|_{t_1}^{t_2}
+ \delta \mcu{F}^{(q)} + \delta \mcu{F}^{(\xi)},\label{eq:mla}\\
\delta \mcu{F}^{(\xi)} \doteq \int^{t_2}_{t_1} \big[\delta \eta \cdot (\dot{\xi} - \pd_\eta H^{(\xi)}) - (\dot{\eta} + \pd_\eta H^{(\xi)}) \cdot \delta \xi \big]\,dt.\notag
\end{gather}
Let us assume that $q(t_{1,2})$ are kept fixed, as in \Sec{sec:coupled}. Then, the requirement $\delta A = 0$ leads to correct equations [\Eqs{eq:ele21} and \eq{eq:he}] if $(q, \xi, \eta)$ are treated as $N + 2M$ independent variables. For this reason, the LAP for our coupled systems can be reformulated as
\begin{gather}
\delta A[q, \xi, \eta] = 0
\end{gather}
with BC still given by \Eqs{eq:dqzero}. (Note that, although $\eta$ is now an independent variable too, it yet does \textit{not} need to be fixed at the end points. Otherwise, the problem would be overdefined.) Accordingly, $A^{\rm (eff)}$ is defined~as
\begin{gather}\label{eq:aeff2}
A^{\rm (eff)}[q] \doteq A[q, \hat{\xi}, \hat{\eta}] + \lambda \cdot \hat{\xi}(t_2),
\end{gather}
where $\lambda$ is a constant Lagrange multiplier. Like earlier, one can show that the solution for $\lambda$ is given by \Eq{eq:aux71}.

%--------------------------------------
\subsection{Symmetrized representation}
\label{sec:sym}

Let us also introduce a ``symmetrized'' representation \cite{foot:mouchet} of the above equations that we will need in \Sec{sec:losc}. Notice that the action \eq{eq:apq} can be rewritten as
\begin{gather}
A[q, \xi, \eta] = \frac{1}{2}\,\eta \cdot \xi\,\big|_{t_1}^{t_2} + \mcu{A}[q, \xi, \eta],\\
\mcu{A}[q, \xi, \eta] \doteq \int^{t_2}_{t_1} \Big[\Lambda^{(q)} 
+ \frac{1}{2}\,(\eta \cdot \dot{\xi} - \dot{\eta} \cdot \xi) - H^{(\xi)}\Big]\,dt.\label{eq:aux33}
\end{gather}
Using $\delta \mcu{A} = \delta A - \delta (\eta \cdot \xi/2)\,\big|_{t_1}^{t_2}$ and \Eq{eq:mla}, one~gets
\begin{gather}
\delta \mcu{A} = \Big[p \cdot \delta q + \frac{1}{2}\,(\eta \cdot \delta \xi - \delta \eta \cdot \xi)\Big]\Big|_{t_1}^{t_2}
+ \delta \mcu{F}^{(q)} + \delta \mcu{F}^{(\xi)}.\label{eq:aux34}
\end{gather}
As usual, we require $\delta q(t_{1,2}) = 0$. Then, we have $M$ BC per end point, and the first term in the square brackets vanishes. The remaining BC can be defined, \eg~as 
\begin{gather}\label{eq:aux102}
(\eta_m \delta \xi^m - \delta \eta_m \xi^m)\,\big|^{t_{1,2}} = 0.
\end{gather}
(Summation over indexes is \textit{not} assumed here; also, the symbol $\smash{\big|^{t_{1,2}}}$ denotes that the equality must be satisfied independently at $t = t_1$ and $t = t_2$.). With obvious reservations, \Eq{eq:aux102} is also equivalent to
\begin{gather}
0 = \delta (\xi^m/\eta_m)\,\big|^{t_{1,2}}.
\end{gather}
This provides precisely $M$ BC per each end point, as needed, and eliminates the second term in the square brackets in \Eq{eq:aux34}. Hence, the requirement of extremal $\mcu{A}$ leads to correct equations [\Eqs{eq:ele21} and \eq{eq:he}]. Therefore, the LAP can be reformulated as
\begin{gather}
\delta \mcu{A}[q, \xi, \eta] = 0
\end{gather}
with BC given by \Eqs{eq:dqzero2} and \eq{eq:aux102}. Since the phase space variables $\xi$ and $\eta$ enter $\mcu{A}$ on the same footing, we call $\mcu{A}$ a phase-space action. 

For $\mcu{A}$, the analog of the constrained variational principle that we introduced in \Sec{sec:constrained} is 
\begin{gather}\label{eq:aux101}
\delta \mcu{A}^{\rm (eff)}[q]  = 0, \quad \delta q(t_{1,2}) = 0,
\end{gather} 
where the action can be taken in the form
\begin{gather}
\mcu{A}^{\rm (eff)}[q] \doteq  \mcu{A}[q, \hat{\xi}, \hat{\eta}] 
+ \sum_m \lambda_m (\hat{\xi}^m/\hat{\eta}_m)\big|^{t_2}.
\end{gather} 
It is easy to see that this VPPD leads to correct equations for $q$ provided that $\lambda_m$ satisfy
\begin{gather}
\lambda_m = - \eta_{m,T}^2,
\end{gather} 
where $\eta_{m,T}$ is the value of $\hat{\eta}_m(t_2)$ corresponding to the physical trajectory for given $q_0$, $q_T$, $\xi_0$, and $\xi'_0$.

Let us also introduce an alternative form of $\mcu{A}^{\rm (eff)}[q]$ that we will need further on. Instead of fixing $\smash{\hat{\xi}^m(t_2)/\hat{\eta}_m(t_2)}$, one can fix $\smash{\hat{\eta}(t_2)}$ and $\smash{\hat{\xi}(t_2)}$ independently at the expense of doubling the number of unknown Lagrange multipliers. Specifically, one can introduce two $M$-dimensional Lagrange multipliers, $\lambda^{(\xi)}$ and $\lambda^{(\eta)}$, and define the effective action as follows:
\begin{gather}\label{eq:aeffsr}
\mcu{A}^{\rm (eff)}[q] \doteq  \mcu{A}[q, \hat{\xi}, \hat{\eta}] 
+ \hat{\eta}(t_2) \cdot \lambda^{(\eta)} + \lambda^{(\xi)} \cdot \hat{\xi}(t_2).
\end{gather} 
Then, one can show that \Eq{eq:aux101} yields correct equations for $q$ provided that
\begin{gather}
\lambda^{(\xi)} = - \eta_T/2, \quad \lambda^{(\eta)} = \xi_T/2,
\end{gather} 
where $\eta_{T}$ and $\xi_T$ are the values of $\hat{\eta}_m(t_2)$ and $\hat{\xi}_m(t_2)$ corresponding to the physical trajectory for given $q_0$, $q_T$, $\xi_0$, and $\xi'_0$. These equations can also be presented in a more compact form, as discussed in \App{app:psp}.

%%%%%%%%%%%%%%%%%%%%%%%%%%%%%%%%%%%%%%%
\section{Coupling to linear oscillators: general theory of linear dispersion}
\label{sec:losc}

Let us now consider, as an example of practical interest, the case when $\mcu{S}^{(\xi)}$ is an ensemble of linear oscillators. In this case, the above phase-space representation facilitates the formulation of a general theory of linear dispersion in a convenient ``quantumlike'' form that is introduced below. Note that truly quantum equations are also subsumed under this formulation. As a side note, the special case of \textit{adiabatic} oscillators can be described by a simpler machinery discussed in \App{app:osc}.

%--------------------------------------
\subsection{Quantumlike formalism}
\label{sec:ql}

Suppose that $\mcu{S}^{(\xi)}$ is an ensemble of linear oscillators. For our purposes, it does not matter what these oscillators are (but see \App{app:landau} for an example); it is sufficient that they can be identified in principle by definition of a linear medium. Then, there exists \cite{my:wkin} a linear transformation of the variables $(\xi, \eta)$ to some variables $(\psi_{\rm r}, \psi_{\rm i})$ (Roman indexes are used to distinguish them from standard italic indexes that denote vector components) such that $\smash{\mcu{A}^{(\xi)}[\psi_{\rm r}, \psi_{\rm i}] = \int_{t_1}^{t_2} \Lambda^{(\xi)}\,dt}$,
\begin{multline}
\Lambda^{(\xi)} = 
\frac{i}{2}\,(\psi^\dag \cdot \dot{\psi} - \dot{\psi}^\dag \cdot \psi)
- \psi^\dag \cdot H_0 \cdot \psi\\
+ \mc{W}(t, \psi, \psi) + \mc{W}^\dag(t, \psi^\dag, \psi^\dag),
\end{multline}
where $\psi \doteq \psi_{\rm r} + i\psi_{\rm i}$ and $\psi^\dag \doteq \transp{\psi_{\rm r}} - i\transp{\psi_{\rm i}}$. (We assume that the oscillators have positive energies for simplicity. Negative-energy oscillations \cite{foot:neg}, which are somewhat exotic and typically unstable yet not impossible, could be included by generalizing the notation as explained in \Ref{my:wkin}.) The matrix $H_0$ is Hermitian and serves as a Hamiltonian. Also, $\mc{W}$ is a bilinear form of $\psi$, and $\mc{W}^\dag$ is the complex-conjugate bilinear form of $\psi^\dag$. The terms $\mc{W}$ and $\mc{W}^\dag$ are determined by the rate at which parameters of the medium evolve in time and vanish when those parameters are time-independent. Also~\cite{my:wkin},
\begin{multline}\label{eq:can}
 (\eta \cdot d\xi - d\eta \cdot \xi)/2 = (i/2)(\psi^\dag \cdot d\psi - d\psi^\dag \cdot \psi)
 \\ =  \psi_{\rm i} \cdot d \psi_{\rm r} - d \psi_{\rm i} \cdot \psi_{\rm r},
\end{multline}
so the BC \eq{eq:aux102} can be expressed as
\begin{multline}\label{eq:bcpsi}
0 = \frac{i}{2}\,(\psi^\dag_m \delta \psi^m - \delta \psi^\dag_m \psi^m)\big|^{t_{1,2}}
 \\ = \big[(\psi_{\rm i})^\dag_m \delta (\psi_{\rm r})^m - (\psi_{\rm r})^\dag_m \delta (\psi_{\rm i})^m\big]\big|^{t_{1,2}},
\end{multline}
or, equivalently, as $\delta [(\psi_{\rm r})^m/(\psi_{\rm i})_m]_{t_{1,2}} = 0$ \cite{foot:contrast}. 

The corresponding ELE are as follows:
\begin{gather}
\frac{\delta \mcu{A}^{(\xi)}}{\delta \psi_{\rm r}} = 0, \quad \frac{\delta \mcu{A}^{(\xi)}}{\delta \psi_{\rm i}} = 0. \label{eq:aux64}
\end{gather}
Yet $\psi_{\rm r}$ and $\psi_{\rm i}$ can be expressed as linear combinations of $\psi$ and $\psi^\dag$. Hence, one can also rewrite \Eqs{eq:aux64} in an equivalent complex form
\begin{gather}
\frac{\delta \mcu{A}^{(\xi)}}{\delta \psi} = 0, \quad \frac{\delta \mcu{A}^{(\xi)}}{\delta \psi^\dag} = 0,
\end{gather}
where $\mcu{A}^{(\xi)}$ is considered a functional of $\psi$ and $\psi^\dag$. In this sense, $\psi$ and $\psi^\dag$ can be formally treated as independent variables. Then, it is easy to see that, under the BC \eq{eq:bcpsi}, one has
\begin{multline}
\delta \mcu{A}^{(\xi)} = \int_{t_1}^{t_2} \big[
\delta \psi^\dag \cdot (i\dot{\psi} - H_0\cdot \psi + \pd \mc{W}^\dag/\pd \psi^\dag) 
\\
+(-i\dot{\psi}^\dag - \psi^\dag \cdot H_0 + \pd \mc{W}/\pd \psi) \cdot \delta \psi
]\,dt.
\end{multline}
Accordingly, the ELE can be cast as follows:
\begin{gather}\label{eq:aux65}
\delta \psi^\dag: \quad i\dot{\psi} = H_0 \cdot \psi - \pd \mc{W}^\dag/\pd \psi^\dag.
\end{gather}
An additional equation is obtained for $\psi^\dag$, but it is simply the adjoint of \Eq{eq:aux65}. (This is understood because we have only two independent real functions, $\psi_{\rm r}$ and $\psi_{\rm i}$, so there can be only one independent complex equation.) Thus, the equation for $\psi^\dag$ will be omitted for brevity.

The terms $\mc{W}$ and $\mc{W}^\dag$ describe effects caused by parametric resonances. (For example, if $\psi$ were the complex probability amplitude of a free quantum particle, they would describe particle annihilation and production.) We ignore such effects for simplicity, so $\mc{W}$ and $\mc{W}^\dag$ are henceforth omitted. Then, the Lagrangian of $\mcu{S}^{(\xi)}$ can be represented simply as follows:
\begin{gather}
\Lambda^{(\xi)} = 
\frac{i}{2}\,(\psi^\dag \cdot \dot{\psi} - \dot{\psi}^\dag \cdot \psi)
- \psi^\dag \cdot H_0 \cdot \psi.
\end{gather}
This leads to a quantumlike equation for $\psi$,
\begin{gather}\label{eq:psih0}
\delta \psi^\dag: \quad i\dot{\psi} = H_0 \cdot \psi,
\end{gather}
which conserves $\psi^\dag \cdot \psi$ even though $H_0$ may depend on time. The quantity $\psi^\dag \cdot \psi$ is understood as the total number of ``quanta'', or ``particles'', in $\smash{\mcu{S}^{(\xi)}}$ \cite{my:wkin}. [Another name of this quantity is the total action of oscillations in $\smash{\mcu{S}^{(\xi)}}$. We prefer not to use this term here to avoid confusion, since $\smash{\mcu{A}^{(\xi)}}$ is also called the action of $\smash{\mcu{S}^{(\xi)}}$.]

%--------------------------------------
\subsection{Interaction model}
\label{sec:co}

Now, let us introduce the interaction between $\mcu{S}^{(\xi)}$ and $\mcu{S}^{(q)}$. We will describe it by an additional Hamiltonian $H^{\rm (int)}$. Assuming that the coupling is sufficiently weak, one can model $H^{\rm (int)}$ with a function linear in $\xi$ and $\eta$. Since $H^{\rm (int)}$ is also real, its representation through $\psi$ and $\psi^\dag$ must then have the form
\begin{gather}\label{eq:hint}
H^{\rm (int)} = - \gamma^\dag \cdot \psi - \psi^\dag \cdot \gamma,
\end{gather}
where $\gamma$ is some complex vector and the minus sign is added for convenience. We will assume $\gamma$ to depend on $(t, q)$ for clarity, but the dependence on derivatives of $q$ could also be included. As a side note, this particular model does not conserve $\psi^\dag \cdot \psi$. A conservative model, which describes parametric interactions and quantum interactions in particular, is discussed in \App{app:q}.

Since $H_0$ in this model can depend only on time, it is convenient to eliminate it using a variable transformation. Specifically, consider a new variable $\Psi$ defined via $\psi \doteq U \cdot \Psi$. Let us choose the operator $U$ (a ``propagator'') such that $i \dot{U} = H_0 \cdot U$ and require that $U$ be a unit matrix at $t = t_1$. This ensures that $U$ remains unitary at all times, so it can be cast as $U = e^{-i\vartheta}$, where $\vartheta$ is some Hermitian operator. More specifically, the propagator can be expressed in terms of the following time-ordered exponential $\msf{T}\exp(\ldots)$:
\begin{gather}\label{eq:mcU}
U(t, t_1) = e^{-i\vartheta} = \msf{T}\,\exp\left[-i \int^t_{t_1} H_0(t')\,dt'\right].
\end{gather}
Also importantly, for any $t_a$, $t_b$, and $t$, one has
\begin{gather}\label{eq:mcU2}
U(t_b, t_a) = U(t_b, t) \cdot U(t, t_a).
\end{gather}

In the new variables, the symmetrized action has the form $\mcu{A}[q, \Psi, \Psi^\dag] = \int_{t_1}^{t_2} L\,dt$, where
\begin{multline}
L = \Lambda^{(q)} (t, q, \dot{q})
+ \frac{i}{2}\,(\Psi^\dag \cdot \dot{\Psi} - \dot{\Psi}^\dag \cdot \Psi) \\
+ \Gamma^\dag(t, q) \cdot \Psi + \Psi^\dag \cdot \Gamma(t, q), \label{eq:LGamma}
\end{multline}
and $\Gamma \doteq U^\dag \cdot \gamma$. The corresponding LAP is formulated~as
\begin{gather}
\delta \mcu{A}[q, \Psi, \Psi^\dag] = 0
\end{gather}
with BC given by \Eq{eq:dqzero2} and
\begin{gather}
\big[(\Psi_{\rm i})^\dag_m \delta (\Psi_{\rm r})^m - (\Psi_{\rm r})^\dag_m \delta (\Psi_{\rm i})^m\big]\big|^{t_{1,2}} = 0.
\end{gather}
It is easy to see that the corresponding ELE are
\begin{align}
\delta q: \quad  & \dot{p} = \pd_q \Lambda^{(q)} + (\pd_q \Gamma^\dag) \cdot \Psi + \Psi^\dag \cdot (\pd_q \Gamma), \label{eq:qfq}\\
\delta \Psi^\dag : \quad & i\dot{\Psi} = - \Gamma.\label{eq:psiQ}
\end{align}

%--------------------------------------
\subsection{Projected dynamics}
\label{sec:pdaeff}

Now suppose that one has been able to express $\Psi$ through $q$. We denote the resulting functional $\hat{\Psi}$ and assume that it is parameterized by some $M$ complex constants (\ie $2M$ real constants), say, $\hat{\Psi}(t_1) = {\Psi}_0$. Hence, we can introduce a VPPD for $\mcu{S}^{(q)}$ as in \Sec{sec:sym} with
\begin{multline}%\notag
\mcu{A}^{\rm (eff)}[q] = \int^{t_2}_{t_1}\Big\{
\Lambda^{(q)} (t, q, \dot{q}) 
\\ + \frac{1}{2}\,[\Gamma^\dag(t, q) \cdot \hat{\Psi} + \hat{\Psi}^\dag \cdot \Gamma(t, q)] \Big\}\,dt + \mc{B},
\end{multline}
\begin{gather}
\mc{B} = \frac{i}{2}\,\big[\nu^\dag \cdot \hat{\Psi}(t_2) - \hat{\Psi}^\dag(t_2) \cdot \nu\big].\label{eq:aux334}
\end{gather}
Here, $\nu$ is a complex Lagrange multiplier introduced such that the action remains real. This representation implies that we seek to fix the values of $\smash{\hat{\Psi}_{\rm r}(t_2)}$ and $\smash{\hat{\Psi}_{\rm i}(t_2)}$ independently from $q(t_2)$ by introducing just the right number of Lagrange multipliers (namely, the $2M$ components of $\nu$, where the real and imaginary parts are counted separately). The values of these multipliers are to be found from the condition that the values of $\smash{\hat{\Psi}_{\rm r}(t_2)}$ and $\smash{\hat{\Psi}_{\rm i}(t_2)}$ must be consistent with the value of $q(t_2)$. It is easily seen that choosing
\begin{gather}\label{eq:mupsi}
\nu = - \Psi_T,
\end{gather}
where $\Psi_T$ is the value of $\hat{\Psi}(t_2)$ corresponding to the physical trajectory for given $\Psi_0$, leads to a correct equation for $q$ [\Eq{eq:qfq}]. Hence, $\mcu{A}^{\rm (eff)}$ is indeed the sought effective action for the projected dynamics of $\mcu{S}^{(q)}$.

%--------------------------------------
\subsection{Polarizability}

%......................................
\subsubsection{Response function}

By integrating \Eq{eq:psiQ}, one can express $\hat{\Psi}$ as follows:
\begin{gather}\label{eq:aux335}
\hat{\Psi} = \Psi_0 + \alpha_\Gamma \circ \Gamma.
\end{gather}
Here, $\alpha_\Gamma$ is a linear integral operator,
\begin{gather}
(\alpha_\Gamma \circ \Gamma)(t) \doteq i \int^t_{t_1} \Gamma[t', q(t')]\,dt',
\end{gather}
which can be understood as the polarizability of $\mcu{S}^{(\xi)}$ with respect to $\Gamma$. (The symbol $\circ$~means ``applied to the expression on the right''.) Hence, the effective Lagrangian can be cast as follows:
\begin{gather}\label{eq:LGq}
\mcu{A}^{\rm (eff)}[q] = \int^{t_2}_{t_1} \Lambda^{(q)}_{0} \,dt + \Xi + \mc{B},\\
\Xi \doteq \text{Re} \int^{t_2}_{t_1} \Gamma^\dag \cdot (\alpha_\Gamma \circ \Gamma)\,dt.
\end{gather}
The difference between $\Lambda^{(q)}_{0} \doteq \Lambda^{(q)} + \text{Re}\,(\Gamma^\dag \cdot \Psi_0)$ and $\Lambda^{(q)}$ produces an effect similar to that of an external force, while $\Xi$ can be described as follows.

Suppose a ``small-amplitude approximation'', namely, that the coordinate $q$ is chosen such that $q = O(a)$, where $a$ is a small parameter. Assuming that the interaction is weak, a typical interaction Hamiltonian will then be linear in $q$, so we assume $\gamma = v \cdot q/\sqrt{2}$, where $v = O(1)$ is an $M \times N$ complex matrix that may depend only on $t$, if at all. (The coefficient $\sqrt{2}$ is introduced to simplify the interpretation of the final result.) This gives
\begin{gather}\label{eq:Gamma}
\Gamma(t, q) = \frac{1}{\sqrt{2}}\,V(t) \cdot q, 
\end{gather}
where $V$ is another $M \times N$ complex matrix, namely,
\begin{gather}\label{eq:Vdef}
V \doteq U^\dag \cdot v = e^{i\vartheta} \cdot v
\end{gather}
(so that, while $v$ is typically slow, $V$ is oscillatory). Then,
\begin{gather}\label{eq:fnPsi}
\hat{\Psi}(t) = \Psi_0 + \frac{i}{\sqrt{2}} \int^{t}_{t_1} V(t') \cdot q(t')\,dt',
\end{gather}
and $\Xi$ can be written as follows:
\begin{gather}
\Xi = \frac{1}{2} \int^{t_2}_{t_1} \transp{q} \cdot (\alpha \circ q)\,dt,
\end{gather}
where $\alpha$ is an integral operator defined as $\alpha \circ q \doteq \text{Re}\,[V^\dag \cdot \alpha_\Gamma \circ (V \cdot q)]$. It is convenient to express this operator in terms of a response (or Green's) function $\alpha$ as
\begin{multline}\label{eq:alphaq}
(\alpha \circ q)(t) = \int^t_{t_1} \alpha(t, t') \cdot q(t')\,dt' \\
= \int^{t_2}_{t_1} \alpha(t, t')\,\Theta(t - t') \cdot q(t')\,dt'.
\end{multline}
Here, $\Theta$ is the Heaviside step function, and $\alpha$ is a real matrix function given by
\begin{align}
\alpha(t, t')  
& \doteq \text{Re} \left[iV^\dag(t) \cdot V(t')\right] \notag \\
& = - \text{Im} \left[V^\dag(t) \cdot V(t')\right] \notag \\
& = (i/2)\,\big[V^\dag(t) \cdot V(t') - V^{\rm T}(t) \cdot V^*(t')\big], \label{eq:akern}
\end{align}
which satisfies 
\begin{gather}\label{eq:attpsym}
\transp{\alpha}(t, t') = - \alpha(t', t).
\end{gather}
Note that ``Re'' and ``Im'' above actually denote the real and imaginary parts, \textit{not} Hermitian and anti-Hermitian parts. In general, $\alpha$ is \textit{not} Hermitian.

As a side remark, the polarizability can be understood as a pseudodifferential operator expressible through $t$ and $i\pd_t$. The fact that the polarizability operator is not necessarily Hermitian can be related to the fact that $i\pd_t$ is not Hermitian on a finite time interval, as opposed to the whole time axis. A related discussion for spatial (rather than temporal) dispersion can be found in \Refs{my:wkin, ref:domingos84}.

Also note that, using \Eq{eq:alphaq}, one can cast $\Xi$~as
\begin{gather}\label{eq:Xittp}
\Xi = \frac{1}{2} \int_{t_1}^{t_2} \int^{t_2}_{t_1} \transp{q}(t) \cdot 
\alpha(t, t')\,\Theta(t - t') \cdot q(t')\,dt\,dt',
\end{gather}
or, equivalently, as
\begin{gather}
\Xi =  \frac{1}{4}\int_{t_1}^{t_2} dt \int_{t_1}^{t_2} dt'\, q^{\rm T}(t) \cdot \alpha(t, t')\, \text{sgn}(t - t')\cdot q(t'),\label{eq:Xisgn}
\end{gather}
where we used \Eq{eq:attpsym} and $\text{sgn}(t) = \Theta(t) - \Theta(-t)$. These formulas will be used below.

%......................................
\subsubsection{Covariant representation, or index lowering}
\label{sec:lower}

If $q$ is understood as a vector, then $\alpha$ serves as a tensor of rank $(1,1)$. Hence, it is convenient to introduce also the related covariant tensor of rank $(0,2)$, \ie ``lower the index'' of the matrix ${\alpha^m}_n$. We denote such index lowering by underlining. Using the corresponding matrix $\underline{\alpha}_{mn}$, one can write
\begin{gather}\label{eq:Gab}
\transp{a} \cdot (\alpha \cdot b) \equiv a_n {\alpha^n}_m b^m = \underline{\alpha}_{nm} b^m a^n = (\underline{\alpha} \cdot b) \cdot a
\end{gather}
for any $a$ and $b$. Accordingly, we can lower the index also in $\alpha$, \ie define
\begin{gather}
(\underline{\alpha} \circ q)(t) \doteq \int^t_{t_1} 
\underline{\alpha}(t, t') \cdot q(t')\,dt'.
\end{gather}
This allows one to write $\transp{(\alpha \circ q)} = \underline{\alpha} \circ q$ (because transposing a vector is the same as lowering its index), and also $\Xi$ can be expressed without involving $\transp{q}$:
\begin{gather}
\Xi = \frac{1}{2} \int^{t_2}_{t_1} (\underline{\alpha} \circ q) \cdot q\,dt.
\end{gather}
If one does not need to distinguish upper and lower indexes, the distinction between $\alpha$ and $\underline{\alpha}$ can be ignored.

%--------------------------------------
\subsection{Derivation of the ELE for $\boldsymbol{q}$}
\label{sec:deriv}

It is instructive to explain at this point how the ELE \eq{eq:qfq} that we anticipate from the general theory can be derived also by a brute-force calculation using
\begin{multline}\label{eq:AeffXiB}
\delta\mcu{A}^{\rm (eff)}[q] = \int_{t_1}^{t_2} \left[\pd_q \Lambda^{(q)} - \dot{p} 
+ \frac{\text{Re}\,(V^\dag \cdot \Psi_0)}{\sqrt{2}}\right] \cdot \delta q\,dt \\ + \delta \Xi + \delta\mc{B}.
\end{multline}

First, using \Eq{eq:Xittp} and the fact that $\transp{a} = a$ for any scalar $a$, one gets
\begin{widetext}
\begin{align}
\delta \Xi 
& = \frac{1}{2} \int_{t_1}^{t_2} \int^{t_2}_{t_1} \left[
\delta \transp{q}(t) \cdot \alpha(t, t')\,\Theta(t - t') \cdot q(t')
+ \transp{q}(t) \cdot \alpha(t, t')\,\Theta(t - t') \cdot \delta q(t')
\right]\,dt\,dt' \notag\\
& = \frac{1}{2} \int_{t_1}^{t_2} \int^{t_2}_{t_1} \left[
\transp{q}(t') \cdot \transp{\alpha}(t, t')\,\Theta(t - t') \cdot \delta q(t)
+ \transp{q}(t) \cdot \alpha(t, t')\,\Theta(t - t') \cdot \delta q(t')
\right]\,dt\,dt' \notag\\
& = \frac{1}{2} \int_{t_1}^{t_2} \int^{t_2}_{t_1} \left[
\transp{q}(t') \cdot \transp{\alpha}(t, t')\,\Theta(t - t') \cdot \delta q(t)
+ \transp{q}(t') \cdot \alpha(t', t)\,\Theta(t' - t) \cdot \delta q(t)
\right]\,dt\,dt' \notag\\
& = \frac{1}{2} \int_{t_1}^{t_2} 
\left[
\int_{t_1}^{t}
\transp{q}(t') \cdot \transp{\alpha}(t, t')\,dt'
+ \int_{t}^{t_2} \transp{q}(t') \cdot \alpha(t', t)\,dt'
\right]
\cdot \delta q(t) \, dt \notag\\
& \equiv \int_{t_1}^{t_2} \Xi^{(q){\rm T}}(t) \cdot \delta q(t)\,dt.\label{eq:xiq}
\end{align}
\end{widetext}
Using \Eq{eq:attpsym}, we can also rewrite $\Xi^{(q)}$ as follows:
\begin{multline}\label{eq:Xiq}
\Xi^{(q)}(t)
= \frac{1}{2} \int_{t_1}^{t} \alpha(t, t') \cdot q(t')\,dt'
\\
- \frac{1}{2} \int_{t}^{t_2} \alpha(t, t') \cdot q(t')\,dt'.
\end{multline}

Next, one gets from \Eq{eq:aux334} that 
\begin{gather}
\delta\mc{B} = \frac{i}{2}\,\big[\nu^\dag \cdot \delta \hat{\Psi}(t_2) - \delta \hat{\Psi}^\dag(t_2) \cdot \nu\big],
\end{gather}
where $\delta \hat{\Psi}(t_2)$ can be obtained from \Eq{eq:fnPsi}; namely,
\begin{gather}
\delta \hat{\Psi}(t_2) = \frac{i}{\sqrt{2}} \int^{t_2}_{t_1} V(t)\cdot \delta q(t)\,dt.
\end{gather}
Let us substitute \Eq{eq:mupsi} and again use \Eq{eq:fnPsi}, now to express $\Psi_T$. This gives
\begin{multline}
\delta\mc{B} = \text{Re} \bigg\{
\left[\Psi_0^\dag - \frac{i}{\sqrt{2}} \int^{t_2}_{t_1} \transp{q}(t')\cdot V^\dag(t')\,dt'\right] \\
\cdot \frac{1}{\sqrt{2}} 
\int^{t_2}_{t_1} V(t)\cdot \delta q(t)\,dt\bigg\},
\end{multline}
or, equivalently,
\begin{gather}
\delta\mc{B} = \int^{t_2}_{t_1} \bigg\{\frac{\text{Re}[\Psi_0^\dag \cdot V(t)]}{\sqrt{2}} + \mc{B}^{(q){\rm T}}\bigg\}\cdot \delta q(t)\,dt,\label{eq:dbq}\\
\mc{B}^{(q)} \doteq \frac{1}{2}\int^{t_2}_{t_1} \alpha(t, t') \cdot q(t')\,dt',
\end{gather}
and we also notice that
\begin{multline}
\mc{B}^{(q)}(t) 
= \frac{1}{2}\int_{t_1}^{t} \alpha(t, t') \cdot q(t') \,dt' \\
 + \frac{1}{2}\int_{t}^{t_2} \alpha(t, t') \cdot q(t') \,dt'.\label{eq:Biq}
\end{multline}

By combining \Eqs{eq:xiq} and \eq{eq:dbq}, we obtain
\begin{multline}
\delta \Xi + \delta \mc{B}= \int_{t_1}^{t_2} \bigg\{\frac{\text{Re}[\Psi_0^\dag \cdot V(t)]}{\sqrt{2}}\\
 + \transp{[\Xi^{(q)}(t) + \mc{B}^{(q)}(t)]}\bigg\} \cdot \delta q(t)\,dt,
\end{multline}
and \Eqs{eq:Xiq} and \eq{eq:Biq} give $\Xi^{(q)} + \mc{B}^{(q)} = \alpha \circ q$. Thus,
\begin{gather}\notag
\delta \Xi + \delta \mc{B}= \int_{t_1}^{t_2} \bigg\{\frac{\text{Re}[\Psi_0^\dag \cdot V(t)]}{\sqrt{2}} 
+ (\underline{\alpha} \circ q)(t) \bigg\}\cdot \delta q(t)\,dt,
\end{gather}
so \Eq{eq:AeffXiB} becomes
\begin{multline}
\delta\mcu{A}^{\rm (eff)}[q] = \int_{t_1}^{t_2} \Big[\pd_q \Lambda^{(q)} - \dot{p} 
+ \underline{\alpha} \circ q  \\
+ \frac{V^\dag \cdot \Psi_0}{\sqrt{2}} + \frac{\Psi_0^\dag \cdot V}{\sqrt{2}} \Big] \cdot \delta q\,dt.
\end{multline}
Then, the requirement $\delta\mcu{A}^{\rm (eff)}[q] = 0$ leads to
\begin{gather}\label{eq:pg}
\delta q: \quad \dot{p} = \pd_q \Lambda^{(q)} + \underline{\alpha} \circ q
+ \frac{V^\dag \cdot \Psi_0}{\sqrt{2}} + \frac{\Psi_0^\dag \cdot V}{\sqrt{2}}.
\end{gather}
Considering the assumed expression for $\Gamma$ [\Eq{eq:Gamma}], this agrees with the general ELE [\Eq{eq:qfq}], as expected.

%%%%%%%%%%%%%%%%%%%%%%%%%%%%%%%%%%%%%%%
\section{Asymptotic approximations}
\label{sec:asym}

In this section, we make approximations to the VPPD and derive reduced theories for certain limiting cases. Although the calculations below may seem cumbersome, producing equivalent results by a brute-force reduction of \Eq{eq:pg} (as opposed to the underlying functional) would still have been harder to do in the general case.

%--------------------------------------
\subsection{Basic concepts}

%......................................
\subsubsection{``Symmetrized'' polarizability}
\label{sec:symm}

Consider a representation of the response function in terms of the ``symmetrized'' time coordinates
\begin{gather}\label{eq:ttau}
\bar{t} \doteq (t + t')/2, \quad \tau \doteq t - t'.
\end{gather}
Specifically, we define 
\begin{gather}
\bar{\alpha}(\bar{t}, \tau) \doteq \alpha(\bar{t} + \tau/2, \bar{t} - \tau/2) 
\equiv \alpha(t, t'),\label{eq:Gdeftau}
\end{gather}
which has the following important property:
\begin{gather}
\bar{\alpha}(\bar{t}, -\tau)= - \transp{\bar{\alpha}}(\bar{t}, \tau).\label{eq:Gsym}
\end{gather}

Using \Eqs{eq:akern}-\eq{eq:Gdeftau} along with \Eq{eq:Vdef}, one can write 
\begin{gather}
\bar{\alpha}(\bar{t}, \tau) = -\text{Im}\,\big[
v^\dag(\bar{t} + \tau/2) \cdot \mc{U}(\bar{t}, \tau) \cdot v(\bar{t} - \tau/2)
\big],\label{eq:avU}\\
\mc{U}(\bar{t}, \tau) \doteq U(\bar{t} + \tau/2, t_1) \cdot U^\dag(\bar{t} - \tau/2, t_1).
\end{gather}
Since $U$ is unitary, one has $U^\dag = U^{-1}$. Using this along with \Eq{eq:mcU2}, we can rewrite $\mc{U}$ as follows:
\begin{align}
\mc{U}(\bar{t}, \tau) 
& = U(\bar{t} + \tau/2, t_1) \cdot U^{-1}(\bar{t} - \tau/2, t_1)\notag\\
& = U(\bar{t} + \tau/2, t_1) \cdot U(t_1, \bar{t} - \tau/2)\notag\\
& = U(\bar{t} + \tau/2, \bar{t} - \tau/2), 
\end{align}
or, in other words,
\begin{gather}\label{eq:mcU3}
\mc{U}(\bar{t}, \tau) = \msf{T}\,\exp\left[-i \int^{\bar{t} + \tau/2}_{\bar{t} - \tau/2} H_0(t')\,dt'\right],
\end{gather}
where $\msf{T}\,\exp(\ldots)$ is a time-ordered exponential.

%......................................
\subsubsection{Phase mixing and the associated small parameters}
\label{sec:pm}

First, consider a stationary medium, \ie a medium where $v$ and $H_0$ are constant. Then, \Eq{eq:mcU3} gives $\mc{U} = \exp(-i H_0 \tau)$, and \Eq{eq:avU} gives
\begin{gather}
\bar{\alpha}(\tau) = - \text{Im}\big(v^\dag \cdot e^{-iH_0\tau} \cdot v\big).
\end{gather}
For simplicity, assume the basis in which $H_0$ is diagonal (``energy basis''); \ie $H_0 = \text{diag}\,(\Omega_1, \Omega_2, \ldots, \Omega_M)$. Then, matrix elements of $\bar{\alpha}$ can be written as
\begin{gather}\label{eq:aux902}
{\bar{\alpha}^m}{}_n(\tau) = -\text{Im}\sum_s{v_s}^{m*} {v^s}_n e^{-i\Omega_s\tau}.
\end{gather}
Provided that the set of $\Omega_s$ is dense enough, the sum over~$s$ can be replaced with an integral [$s \mapsto s(\Omega)$]:
\begin{gather}
\bar{\alpha}(\tau) \approx - \text{Im} \int e^{-i\Omega\tau} f(\Omega)\,d\Omega,
\end{gather}
where ${f^m}{}_n(\Omega) \doteq (ds/d\Omega)\,[{v_{s(\Omega)}}^{m*} {v^{s(\Omega)}}_n]$. Since $\bar{\alpha}(\tau)$ is thereby a Fourier image of $f(\Omega)$ (or vice versa), it approaches zero at $\tau \to \infty$, at least if $f(\Omega)$ is continuous. If the characteristic gap $\Delta \Omega$ between neighboring eigenfrequencies is nonzero (and $M \gg 1$), the same conclusion applies at $\tau \ll \Delta \Omega^{-1}$, when the spectrum discreteness is inessential. Thus, as long as larger $\tau$ are not of interest (as will be assumed below), there is no need to distinguish continuous and discrete spectra.

The effect $\bar{\alpha}(\tau \to \infty) \to 0$ is known as phase mixing. Although introduced here for a stationary medium, it extends also to media evolving at large enough time scales $\mc{T}$, namely,
\begin{gather}\label{eq:vro}
\frac{\tau_{\rm pm}}{\mc{T}} \ll 1, \quad 
\bar{\Omega} \mc{T}\,\left(\frac{\tau_{\rm pm}}{\mc{T}}\right)^2 \ll 1,
\end{gather}
where $\bar{\Omega}$ is the characteristic eigenvalue of $H_0$. The former requirement ensures that $v$ and $H_0$ be approximately constant on the time scale $\tau_{\rm pm}$. The latter requirement allows approximating \Eq{eq:mcU3} much like in the case of a stationary medium: 
\begin{gather}\label{eq:mcUasym}
\mc{U}(\tau, \bar{t}) \approx \exp[-i H_0(\bar{t}) \tau].
\end{gather}
Our intention is to discuss approximations to the VPPD that become possible in this limit. In order to do that, let us introduce some definitions first.

%......................................
\subsubsection{Polarizability in the spectral representation}
\label{sec:weyl}

Let us introduce an auxiliary matrix function
\begin{gather}
\widetilde{\alpha}_\omega(\bar{t}) \doteq \int^{\tau_m}_0 \bar{\alpha}(\bar{t}, \tau)\,e^{i\omega \tau}\,d\tau,
\end{gather}
where $\omega$ is assumed real,  $\bar{\alpha}$ is given by \Eq{eq:Gdeftau}, and $\tau_m$ is some constant. (In later sections, we will treat $\tau_m$ as a function of $\bar{t}$, but this distinction is not important here.) Let us also introduce the Hermitian and anti-Hermitian parts of $\widetilde{\alpha}_\omega$, namely,
\begin{gather}\label{eq:alphaHA}
\widetilde{\alpha}_{\omega, H} \doteq \frac{\widetilde{\alpha}_{\omega}+\widetilde{\alpha}_{\omega}^\dag}{2},
\quad
\widetilde{\alpha}_{\omega, A} \doteq \frac{\widetilde{\alpha}_{\omega}-\widetilde{\alpha}_{\omega}^\dag}{2i},
\end{gather}
such that $\widetilde{\alpha}_{\omega}=\widetilde{\alpha}_{\omega, H}+i \widetilde{\alpha}_{\omega, A}$. (In this notation, both $\widetilde{\alpha}_{\omega, H}$ and $\widetilde{\alpha}_{\omega, A}$ are Hermitian, so $i\widetilde{\alpha}_{\omega, H}$ and $i\widetilde{\alpha}_{\omega, A}$ are anti-Hermitian.) As shown in \App{app:alphas},
\begin{gather}
\widetilde{\alpha}_{\omega, H}(\bar{t})=\frac{1}{2}\int_{-\tau_m}^{\tau_m} \bar{\alpha}(\bar{t}, \tau)\,\text{sgn}(\tau)\,e^{i\omega\tau}\,d\tau,\label{eq:aproph}\\
\widetilde{\alpha}_{\omega, A}(\bar{t})=\frac{1}{2i}\int_{-\tau_m}^{\tau_m} \bar{\alpha}(\bar{t}, \tau)\,e^{i\omega\tau}\,d\tau.\label{eq:apropa}
\end{gather}

Provided efficient phase mixing, the function $\widetilde{\alpha}_\omega$ has a well-defined limit at large $\tau_m$. We call this limit $\alpha_{\omega}$; \ie
\begin{multline}\label{eq:aw}
\alpha_{\omega}(\bar{t})\doteq \int_0^{\infty}\bar{\alpha}(\bar{t}, \tau)\,e^{i \omega\tau}\,d\tau\\
\equiv \int_{-\infty}^\infty \alpha(\bar{t} + \tau/2, \bar{t} - \tau/2)\,\Theta(\tau)\,e^{i \omega \tau}\,d\tau.
\end{multline}
The function $\alpha_{\omega}$ is recognized as the Laplace image of $\bar{\alpha}$ evaluated at real $\omega$. (Of course, one can also calculate it by taking the Laplace transform of $\bar{\alpha}$ for frequencies with $\text{Im}\,\omega > 0$ and then taking the limit $\text{Im}\,\omega \to 0$; then $\tau_m$ does not need to be introduced at all. We use this method in \App{app:landau}, where a specific example is considered.) Even more generally, the function $\alpha_{\omega}$ can be understood as the \textit{Weyl image of the polarizability operator} \cite{foot:weyl}. Like for $\widetilde{\alpha}_\omega$, the Hermitian and anti-Hermitian parts of $\alpha_\omega$~
are
\begin{gather}
\alpha_{\omega, H}(\bar{t})=\frac{1}{2}\int_{-\infty}^{\infty} \bar{\alpha}(\bar{t}, \tau)\,\text{sgn}(\tau)\,e^{i\omega\tau}\,d\tau,\\
\alpha_{\omega, A}(\bar{t})=\frac{1}{2i}\int_{-\infty}^{\infty} \bar{\alpha}(\bar{t}, \tau)\,e^{i\omega\tau}\,d\tau.
\end{gather}
Approximate formulas for $\widetilde{\alpha}_\omega$ and $\alpha_\omega$ in the limit \eq{eq:vro} can be found in \App{app:alphas}. It is also shown there that $\alpha_{\omega, H}$ and $\alpha_{\omega, A}$ satisfy the Kramers-Kronig relations.

%--------------------------------------
\subsection{Quasistatic limit}
\label{sec:qs}

First, let us consider the case when $q$ and, possibly, the medium evolve on time scales $\mc{T}$ that satisfy
\begin{gather}
\epsilon_0 \doteq \text{max}\left\{
\frac{\tau_{\rm pm}}{\mc{T}},\,
\bar{\Omega} \mc{T} \left(\frac{\tau_{\rm pm}}{\mc{T}}\right)^2,
\frac{1}{\bar{\Omega} \mc{T}}
\right\} \ll 1.
\end{gather}
(To simplify the notation, we will assume all the three small parameters to be comparable to each other, but this is not essential.) Below, we derive an asymptotic expression for $\mcu{A}^{\rm (eff)}$ using $\epsilon_0$ as a small parameter.

\begin{figure*}
\centering
\includegraphics[width=.9\textwidth]{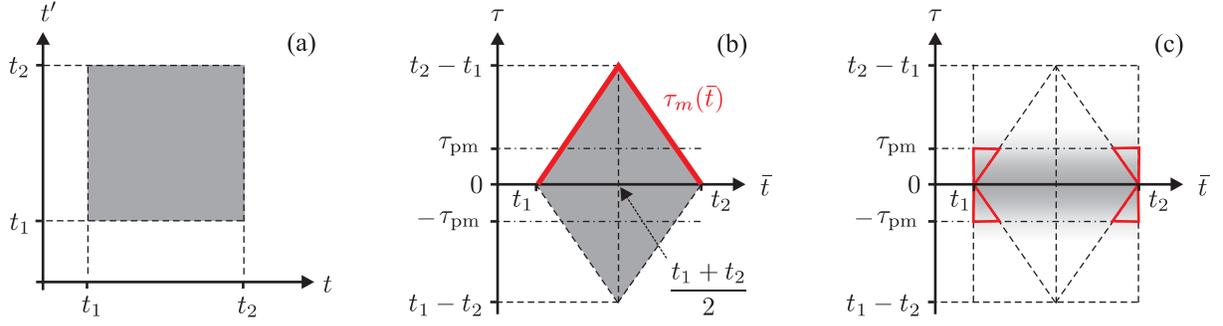}
\caption{Integration domains (shaded): (a) The exact domain in $(t, t')$ space; namely, $(t, t') \in (t_1, t_2) \times (t_1, t_2)$. (b) The exact domain in $(\bar{t}, \tau)$ space; namely, $(\bar{t}, \tau) \in (t_1, t_2) \times (-\tau_m(\bar{t}), \tau_m(\bar{t}))$. The function $\tau_m(\bar{t})$ is plotted in bold red. (c) The approximate domain in $(\bar{t}, \tau)$ space; namely, $(\bar{t}, \tau) \in (t_1, t_2) \times (-\infty, \infty)$. This approximation is justified by the fact that the integrand is negligible at $|\tau| \gtrsim \tau_{\rm pm}$, as indicated by a gradient fill. The red triangles indicate the error introduced by this approximation.}
\label{fig:domain}
\end{figure*}  

%......................................
\subsubsection{Expression for $\Xi$}

To approximate the term $\Xi$ as given by \Eq{eq:Xisgn}, let us first map the integration variables $(t, t')$ in the double integral to $(\bar{t}, \tau)$ using \Eqs{eq:ttau}. The integration domain is then mapped as shown in Figs.~\ref{fig:domain}(a) and (b), namely, to $(\bar{t}, \tau) \in (t_1, t_2) \times (-\tau_m(\bar{t}), \tau_m(\bar{t}))$, where $\tau_m(\bar{t})$ is a piecewise-linear function. This gives
\begin{gather}
\Xi = \frac{1}{4}\,\text{Tr}\int_{t_1}^{t_2} d\bar{t} \int^{\tau_m(\bar{t})}_{-\tau_m(\bar{t})} d\tau\, w(\bar{t}, \tau) \cdot \bar{\alpha}(\bar{t}, \tau)\,\text{sgn}(\tau),
\end{gather}
where Tr stands for ``trace'', and $w$ is a matrix given by
\begin{gather}\notag
w(\bar{t}, \tau) \doteq q(\bar{t} - \tau/2)\,\transp{q}(\bar{t} + \tau/2) = q(\bar{t})\,\transp{q}(\bar{t}) + O(\epsilon_0).
\end{gather}
To the lowest (zeroth) order in $\epsilon_0$, one has
\begin{multline}
\Xi \approx \frac{1}{4}\,\text{Tr}\int_{t_1}^{t_2} d\bar{t}\,w(\bar{t},0) \cdot \int^{\tau_m(\bar{t})}_{-\tau_m(\bar{t})} d\tau\, \bar{\alpha}(\bar{t}, \tau)\,\text{sgn}(\tau)\\
= \frac{1}{2} \int_{t_1}^{t_2} \transp{q}(\bar{t}) \cdot \widetilde{\alpha}_{0,H}(\bar{t}) \cdot q(\bar{t})\,d\bar{t}.
\end{multline}
The function $\widetilde{\alpha}_{0,H}$ [given by \Eq{eq:aproph} at $\omega = 0$] can be approximated with $\alpha_{0, H}$ everywhere except for an $O(\epsilon_0)$ part of the integration domain. Neglecting this $O(\epsilon_0)$ correction, we get
\begin{multline}
\Xi \approx \frac{1}{2} \int_{t_1}^{t_2} \transp{q}(\bar{t}) \cdot \alpha_0(\bar{t}) \cdot q(\bar{t})\,d\bar{t}
\\ \equiv
\int_{t_1}^{t_2} [\underline{\alpha}_0(\bar{t}) \cdot q(\bar{t})] \cdot q(\bar{t})\,d\bar{t},
\end{multline}
which corresponds to the approximation illustrated in Fig.~\ref{fig:domain}(c). We have omitted the index $H$ in $\alpha_{0, H}$ because, in the zero-frequency limit, the polarizability's Weyl symbol \eq{eq:aw} becomes Hermitian (\App{app:alphas}); \ie $\alpha_0 \approx \alpha_{0,H}$. Specifically,
\begin{gather} \label{eq:alpha0est}
\alpha_0 \approx v^\dag \cdot H_0^{-1} \cdot v
\end{gather}
as seen from \Eq{eq:zerow} in the limit when the oscillating term is eliminated by phase mixing. 

%......................................
\subsubsection{Expression for $\mc{B}$}

Using \Eq{eq:fnPsi}, let us express $\hat{\Psi}$ as
\begin{gather}\label{eq:intpsi0}
\hat{\Psi} = \Psi_0 + \frac{i}{\sqrt{2}} \int^t_{t_1} e^{i\vartheta(t')} \cdot v(t') \cdot q(t')\,dt'.
\end{gather}
The function $\vartheta$ is such that, at zero $\epsilon_0$, it has a constant derivative equal to $H_0$. Hence, at small nonvanishing $\epsilon_0$, one can expect $\pd_t \vartheta$ to be slow and close to $H_0$. Then, the integrand in \Eq{eq:intpsi0} is a rapidly oscillating exponent times a slow function $v \cdot q$. Such integral is entirely determined by values of the integrand at the ends of the integration domain and is of order $||v||\,||q||/\bar{\Omega}$ \cite{foot:adiab}. This gives the following estimate for the term $\mc{B}$ in \Eq{eq:aux334}:
\begin{gather}\label{eq:Best}
\mc{B} \sim ||v||^2\,||q||^2/\bar{\Omega}^2,
\end{gather}
assuming the effect of $\Psi_0$ vanishes due to phase mixing. Hence, $\mc{B}/\Xi \sim \epsilon_0$.

%......................................
\subsubsection{Final result}

By combining the above results and neglecting corrections $O(\epsilon_0)$ and smaller, one obtains
\begin{gather}
\mcu{A}^{\rm (eff)}[q] = \int^{t_2}_{t_1} [\Lambda^{(q)}_{0} (t, q, \dot{q}) - \Phi_0(t, q)]\, dt,\\
\Phi_0(t, q) \doteq - \frac{1}{2}\,[\underline{\alpha}_0(t) \cdot q] \cdot q.\label{eq:Udef}
\end{gather}
Clearly, the corresponding ELE is as follows:
\begin{gather}
\delta q: \quad \dot{p} = \pd_q \Lambda^{(q)}_{0} + \underline{\alpha}_0 \cdot q.
\end{gather}
These results indicate that the interaction of a quasistatic system with a linear medium $\mcu{S}^{(\xi)}$ is equivalent to the interaction with an effective potential $\Phi_0$ given by \Eq{eq:Udef}. The symmetric matrix $\underline{\alpha}_0$ represents the covariant form of~$\alpha_0$ given by \Eq{eq:alpha0est} and serves as the polarizability of $\mcu{S}^{(\xi)}$ with respect to $q$. Also note that, for positive-definite Hamiltonian $H_0$, the polarizability $\alpha_0$ is positive-definite too, and thus $\Phi_0 < 0$.

%--------------------------------------
\subsection{Limit of fast oscillations}
\label{sec:fastglobal}

Here, we derive an asymptotic variational principle for the case when $q$ oscillates rapidly compared to the evolution of the medium. The method is akin to Whitham's averaging \cite{book:whitham}, but the averaging procedure is justified differently. Specifically, this is done as follows.

%......................................
\subsubsection{Basic equations}
\label{sec:gobasic}

Suppose $\Lambda^{(q)}$ is bilinear in $(q, \dot{q})$. Then, \Eq{eq:pg} is linear and allows complex solutions. Let us consider a complex solution $q_+$ of the form $q_+ = Q(t)\,e^{i \theta(t)}$, where $Q$ is a complex vector and $\theta$ is a real phase. 
Provided that the time scale $\mc{T}$ at which the medium evolves is large enough, in principle it is possible to construct asymptotic $Q$ and $\theta$ such that $\smash{\omega (t)\doteq -\dot{\theta}(t)}$ is slow compared to $\theta$ at all $t$; \ie $q$ is quasimonochromatic. (The effect of $\Psi_0$ will be assumed negligible, like in \Sec{sec:qs}.) More specifically, harmonics of $\omega$ remain suppressed with exponential accuracy in the small parameter $(\omega \mc{T})^{-1}$. (If the system supports more than one mode, we assume that only one mode is excited, and the spectral distance to other modes is $\Delta \omega \gtrsim \omega$.) Here, we assume that $\mc{T}$ is also the time scale of the evolution of $Q$ and $\omega$, and $\mc{T} \sim T$.

Clearly, if $q_+ = Q(t)\,e^{i \theta(t)}$ is a solution, then $q_- = Q^*(t)\,e^{-i \theta(t)}$ a solution too, and so is $(q_+ + q_-)/2$. Since the latter is also real, it can be adopted as an \textit{exponentially accurate asymptotic model} of a realizable physical process, which we can also express as follows:
\begin{gather}
q = \text{Re}\,[Q(t)\,e^{i \theta (t)}].
\end{gather}
Let us substitute this model into \Eq{eq:LGamma}. As usual \cite{foot:adiab}, the rapidly oscillating terms that couple $q_+$ and $q_-$ can be eliminated, which is also expected from the definition of $q_+$ and $q_-$. (Such ``$\theta$-averaging'' is denoted below with angular brackets, $\favr{\ldots}$.) This implies that the medium responds to $q_+$ and $q_-$ independently, as if there were two independent subsystems $\mcu{S}^{(\xi)}$ that interact with $q_+$ and $q_-$, respectively. The corresponding responses will be denoted $\Psi_{\pm}$. Hence, instead of the Lagrange multipliers $\nu$ and $\nu^\dag$  that we introduced before, now we must introduce twice as much Lagrange multipliers. We denote them $\nu_{\pm}$ and $\nu_{\pm}^\dag$. Hence, like in \Sec{sec:pdaeff}, the effective action \eq{eq:LGq} can be represented as follows:
\begin{gather}
\mcu{A}^{\rm (eff)} = \favr{A^{(q)}} + \favr{\Xi} + \favr{\mc{B}}, \label{eq:aeffgo1}\\
\favr{A^{(q)}} = \int_{t_1}^{t_2} \favr{\Lambda^{(q)}}\,dt,
\end{gather}
\begin{multline}
\favr{\Xi} = \frac{1}{8} \int_{t_1}^{t_2} dt \int_{t_1}^{t_2} dt'\,e^{i\theta(t') - i \theta(t)}
\\\times Q^\dag(t) \cdot \alpha(t, t')\,\text{sgn} (\tau) \cdot Q(t')\label{eq:XiQsgn}
\end{multline}
[where we used \Eq{eq:Xisgn}], and $\favr{\mc{B}} =  \mc{B}_+ + \mc{B}_-$, where
\begin{gather}\label{eq:auxB}
\mc{B}_{\pm} = - \frac{1}{\sqrt{2}}\,\text{Re}\sum_{\pm}\left[\nu_{\pm}^\dag \cdot \int_{t_1}^{t_2} V(t')\cdot Q_\pm(t') \,e^{\pm i\theta(t')} \,dt'\right],
\end{gather}
assuming the notation $Q_+ \doteq Q$ and $Q_- \doteq Q^*$. Also like in \Sec{sec:pdaeff}, one finds
\begin{gather}\label{eq:nupm}
\nu_{\pm} = - \frac{\Psi_{\pm,T}}{4} = -\frac{i}{4\sqrt{2}} \int_{t_1}^{t_2} V(t)\cdot Q_\pm(t) \,e^{\pm i\theta(t)}\,dt.
\end{gather}
[The quantity $\nu_{\pm}$ must not be confused with $\nu$ given by \Eq{eq:mupsi}. In particular, note the extra factor $1/4$.]

As usual, the functions $\theta$, $Q$, and $Q^\dag$ are considered independent. Then,
\begin{multline}\notag
\delta \mcu{A}^{\rm (eff)} = \int_{t_1}^{t_2}
\bigg[
\frac{\delta \mcu{A}^{\rm (eff)}}{\delta \theta}\, \delta \theta 
\\ + \sum_{n=1}^{\infty} 
\left(\frac{\delta \mcu{A}^{\rm (eff)}}{\delta Q} \cdot \delta Q + \delta Q^\dag \cdot \frac{\delta \mcu{A}^{\rm (eff)}}{\delta Q^\dag}\right)
 \bigg]\,dt.
\end{multline}
We seek to approximate $\delta \mcu{A}^{\rm (eff)}$ to the zeroth order in the small parameter
\begin{gather}\label{eq:epsilonom}
\epsilon_\omega \doteq \max \left\{
\frac{\tau_{\rm pm}}{\mc{T}},\,
\omega \mc{T}\left(\frac{\tau_{\rm pm}}{\mc{T}}\right)^2,
\frac{1}{\omega \mc{T}}
\right\}\ll 1,
\end{gather}
where $\tau_{\rm pm}$ is the phase-mixing time (\Sec{sec:pm}). (As in \Sec{sec:qs}, we adopt for clarity that all the three parameters are comparable to each other.) We assume $\bar{\Omega} \sim \omega$ and also the standard ordering for $\alpha_{\omega}$ (\Sec{sec:weyl}), which will be be verified \textit{a~posteriori}:
\begin{gather}\label{eq:alphaord}
\alpha_{\omega, H} = O(1),
\quad
\alpha_{\omega, A} \lesssim O(\epsilon_\omega).
\end{gather}
[It is to be noted that the assumptions \eq{eq:epsilonom} and \eq{eq:alphaord} are standard in GO theory of electromagnetic waves \cite{book:kravtsov}.] Then, the individual terms on the right side of \Eq{eq:aeffgo1} can be calculated as follows.

%......................................
\subsubsection{Approximation of $\favr{A^{(q)}}$} 
\label{sec:Aq}

Since $\Lambda^{(q)}$ is a local function of $(q, \dot{q})$, the function $\favr{\Lambda^{(q)}}$ cannot depend on $\theta$ other than through $\omega$. Hence, 
\begin{multline}\notag
\delta \favr{A^{(q)}} = \int_{t_1}^{t_2} \bigg[ \frac{\pd}{\pd t}\left(\frac{\delta \favr{\Lambda^{(q)}}}{\delta \omega}\right) \delta \theta
\\ + \frac{\delta \favr{\Lambda^{(q)}}}{\delta Q} \cdot \delta Q 
+ \delta Q^\dag \cdot \frac{\delta \favr{\Lambda^{(q)}}}{\delta Q^\dag}\bigg]\, dt,
\end{multline}
where we used $\delta \omega = -\pd_t (\delta \theta)$ and integrated by parts with $\delta\theta(t_{1,2}) = 0$. Since
\begin{gather}\label{eq:dgosc}
\delta \theta = O(\epsilon_\omega^{-1}), \quad \delta Q = O(1),
\end{gather}
while $\smash{\pd_t(\delta_\omega \favr{\Lambda^{(q)}}) = O(\epsilon_\omega)}$, $\smash{\delta_Q \favr{\Lambda^{(q)}} = O(1)}$, and $\smash{\delta_{Q^\dag} \favr{\Lambda^{(q)}} = O(1)}$, all the three terms in the square brackets are $O(1)$. Hence, within the accuracy of our model, it is enough to approximate $\favr{A^{(q)}}$ to the zeroth order in $\epsilon_\omega$. Since $\Lambda^{(q)}$ was assumed to be a bilinear function of $(q, \dot{q})$, the functional $\favr{A^{(q)}}$ then has the following form:
\begin{gather}\label{eq:AD}
\favr{A^{(q)}} \approx \frac{1}{4} \int_{t_1}^{t_2} \text{Re}\,[Q^\dag(t)\cdot \msf{D}_{\omega(t)}^{(q)}(t) \cdot Q(t)]\,dt,
\end{gather}
where the factor $1/4$ is introduced for convenience, and $\smash{\msf{D}_{\omega(t)}^{(q)}}$ is some matrix that is determined by the local frequency $\omega(t)$ and, in nonstationary systems, also $t$. [Below, we also refer to this matrix using a shortened notation $\smash{\msf{D}_\omega^{(q)}}$.] The anti-Hermitian part of this matrix provides zero contribution to $\smash{\text{Re}\,[Q^\dag \cdot \msf{D}_\omega^{(q)} \cdot Q]}$, so we can assume that $\smash{\msf{D}_\omega^{(q)}}$ is Hermitian for all real $\omega$ without loss of generality. Then, ``Re'' in \Eq{eq:AD} can be omitted. 

%......................................
\subsubsection{Approximation of $\favr{\Xi}$, dressed action} 

Let us now approximate $\favr{\Xi}$ given by \Eq{eq:XiQsgn}. First, note that, due to \Eq{eq:ttau}, we have
\begin{gather}
\theta(t') = \theta(\bar{t}) - \frac{\tau}{2}\,\pd_t \theta(\bar{t}) + \frac{\tau^2}{8}\,\pd_t^2 \theta(\bar{t}) - \frac{\tau^3}{48}\,\pd_t^3 \theta(\bar{t}) + \ldots, \notag\\
\theta(t) = \theta(\bar{t}) + \frac{\tau}{2}\,\pd_t \theta(\bar{t}) + \frac{\tau^2}{8}\,\pd_t^2 \theta(\bar{t}) + \frac{\tau^3}{48}\,\pd_t^3 \theta(\bar{t}) + \ldots. \notag
\end{gather}
Hence,
\begin{multline}
\theta(t') - \theta(t) = - \tau \,\pd_t \theta(\bar{t}) - \frac{\tau^3}{24}\,\pd_t^3 \theta(\bar{t}) + \ldots\\
= \tau \omega(\bar{t}) + \frac{\tau^3}{24}\,\pd_t^2 \omega(\bar{t}) + \ldots,
\end{multline}
where the second term on the right is estimated as
\begin{gather}
\tau^3 \pd_t^2 \omega \sim \frac{\tau_{\rm pm}}{\mc{T}}\,\left[
\omega \mc{T}\,\left(\frac{\tau_{\rm pm}}{\mc{T}}\right)^2
\right] = O(\epsilon_\omega^2) \ll 1.
\end{gather}
This gives
\begin{gather}
e^{i\theta(t') - i\theta(t)} = e^{i\omega(\bar{t})\tau  + O(\epsilon_\omega^2)} = e^{i\omega(\bar{t})\tau} + O(\epsilon_\omega^2).
\end{gather}
Omitting $O(\epsilon_\omega^2)$, one can rewrite \Eq{eq:XiQsgn} as follows:
\begin{multline}
\favr{\Xi} \approx \frac{1}{8}\,\text{Re}\,\text{Tr}\int_{t_1}^{t_2} dt \int_{t_1}^{t_2} dt'\,W(\bar{t}, \tau)\\
\cdot \bar{\alpha}(\bar{t}, \tau)\,\text{sgn}(\tau)\,e^{i\omega(\bar{t})\tau}.\label{eq:Xiaux}
\end{multline}
Here, the matrix $W$ is given by
\begin{gather}\label{eq:WW}
W(\bar{t}, \tau)\doteq Q(t')\, Q^\dag (t) 
= Q(\bar{t}) \,Q^\dag(\bar{t}) + O(\epsilon_\omega).
\end{gather}
Since the integrand in \Eq{eq:Xiaux} does not depend on $\theta$ explicitly, the correction $O(\epsilon_\omega)$ can be omitted for the same reason as in \Sec{sec:Aq}. Hence, we switch from integration in $(t, t')$ to the integration in $(\bar{t}, \tau)$ and get
\begin{gather}
\favr{\Xi} 
\approx \frac{1}{4}\,\text{Tr}\int_{t_1}^{t_2} W(\bar{t}, 0) \cdot \widetilde{\alpha}_{\omega, H}(\bar{t})\,d\bar{t}.
\end{gather}
Note that $\widetilde{\alpha}_{\omega, H} \sim \alpha_{\omega, H} = O(1)$ (\Sec{sec:gobasic}). This function can be approximated with $\alpha_{\omega, H}$ everywhere except for an $O(\epsilon_\omega)$ part of the integration domain. Since corrections $O(\epsilon_\omega)$ in $\favr{\Xi}$ are deemed negligible, we obtain
\begin{gather}\label{eq:XiXi}
\favr{\Xi} \approx \frac{1}{4}\int_{t_1}^{t_2} Q^\dag \cdot \alpha_{\omega, H} \cdot Q \,dt.
\end{gather}

As a side note, one can also rewrite \Eq{eq:XiXi} as $\favr{\Xi} \approx \smash{-\int_{t_1}^{t_2}\Phi_\omega(t)\,dt}$, so $\favr{\Xi}$ is understood as a contribution of an effective potential
\begin{gather}
\Phi_\omega\doteq -\frac{1}{4}\,Q^\dag \cdot \alpha_{\omega, H} \cdot Q
\end{gather}
that the subsystem $\mcu{S}^{(q)}$ experiences due to the interaction with the subsystem $\mcu{S}^{(\xi)}$. The potential $\Phi_\omega$ represents the ponderomotive energy (or, more precisely, the ponderomotive term in the Lagrangian) of $\mcu{S}^{(\xi)}$. In the context of particle dynamics with Hermitian $\alpha_\omega$, the obtained linear relation between $\Phi_\omega$ and $\alpha_\omega$ is also known as the $K$-$\chi$ theorem \cite{my:kchi, ref:kaufman87, ref:cary77, tex:myqponder}.

Since $\favr{\Xi}$ has the same form as $\favr{A^{(q)}}$ [\Eq{eq:AD}], let us combine them into a ``dressed action'' $\smash{A_{\text{dr}}^{(q)}} \doteq \smash{\favr{A^{(q)}} +\favr{\Xi}}$. For that, let us introduce $\smash{\msf{D}_{\omega} \doteq \msf{D}^{(q)}_\omega + \alpha_{\omega}}$, whose Hermitian and anti-Hermitian parts are
\begin{gather}
\msf{D}_{\omega, H} = \msf{D}^{(q)}_\omega + \alpha_{\omega, H},
\quad
\msf{D}_{\omega, A} = \alpha_{\omega, A}.
\end{gather}
Then, 
\begin{gather}
A_{\text{dr}}^{(q)} = \frac{1}{4}\int_{t_1}^{t_2} Q^\dag \cdot \msf{D}_{\omega, H}\cdot Q \, dt.\label{eq:Adr}
\end{gather}
This can also be written as $A_{\text{dr}}^{(q)}=\int_{t_1}^{t_2} \mc{L}\,dt$, where
\begin{gather}
\mc{L}(t, \omega, Q, Q^\dag) \doteq \frac{1}{4}\,Q^\dag \cdot \msf{D}_{\omega, H}(t)\cdot Q
\end{gather}
is understood as the dressed Lagrangian of $\mcu{S}^{(q)}$. Accordingly, one gets
\begin{multline}\label{eq:dadr}
\delta A_{\text{dr}}^{(q)} = \int_{t_1}^{t_2} \bigg[(\pd_t I)\, \delta \theta 
+ \frac{1}{4}\, \delta Q^\dag \cdot \left(\msf{D}_{\omega, H}\cdot Q\right)
\\+ \frac{1}{4}\,(Q^\dag \cdot \msf{D}_{\omega, H}) \cdot \delta Q\bigg]\, dt,
\end{multline}
where
\begin{gather}
I \doteq \pd_{\omega}\mc{L}(t, \omega, Q, Q^\dag) 
= Q^\dag \cdot \pd_{\omega} \msf{D}_{\omega, H} \cdot Q/4
\end{gather}
is understood as the number of quanta of the oscillation mode, or the action of this mode \cite{my:amc, my:wkin} [again, not to be confused with the action such as $\mcu{A}^{\rm (eff)}$].

%......................................
\subsubsection{Approximation of $\delta \mc{B}_\pm$}
\label{app:goauxB}

A straightforward calculation (\App{app:dB}) gives
\begin{gather}
\delta \favr{\mc{B}} = \delta \mc{B}_\theta + \delta \mc{B}_Q + O(\epsilon_\omega^2)\label{eq:aux3001},\\
\delta \mc{B}_\theta \doteq -\frac{i}{4}\,\text{Tr} \int_{t_1}^{t_2}dt \int_{t_1}^{t_2} dt'\, e^{i\omega \tau} \,
\,\delta \theta(\bar{t})\,W(\bar{t}, \tau)\cdot \bar{\alpha}(\bar{t}, \tau),\label{eq:dBtheta}\\
\delta \mc{B}_Q \doteq -\frac{i}{4}\,\text{Tr} \int_{t_1}^{t_2}dt \int_{t_1}^{t_2} dt'\, e^{i\omega \tau} \,
\delta Y(\bar{t}, \tau) \cdot \bar{\alpha}(\bar{t}, \tau),\label{eq:dBQ}\\
\delta Y(\bar{t}, \tau) \doteq \frac{i}{2}
\left[Q(t')\cdot \delta Q^\dag (t)-\delta Q(t')\cdot Q^\dag (t)\right].\label{eq:dY}
\end{gather}
Using $W(\bar{t}, \tau) \approx W(\bar{t}, 0) + \tau \,\pd_\tau W(\bar{t}, 0)$, one further gets
\begin{multline}
\delta \mc{B}_\theta = \frac{1}{2}\,\text{Tr} \int_{t_1}^{t_2} d\bar{t}\,\delta \theta(\bar{t})
\\ \times \big\{W(\bar{t}, 0) - i [\pd_\tau W(\bar{t}, 0)] \pd_\omega \big\} \cdot \widetilde{\alpha}_{\omega, A}.
\end{multline}
Note that $\widetilde{\alpha}_{\omega, A} \sim \alpha_{\omega, A} = O(\epsilon_\omega)$ (\Sec{sec:gobasic}) and $\delta \theta = O(\epsilon_\omega^{-1})$ [\Eq{eq:dgosc}]. Since we need to calculate $\delta \mc{B}_\theta$ to the zeroth order in $\epsilon_\omega$, the contribution of $\pd_\tau W \sim W/\mc{T}$ can then be neglected. Hence, 
\begin{gather}
\delta \mc{B}_\theta \approx \frac{1}{2}\,\int_{t_1}^{t_2} [Q^\dag(\bar{t}) \cdot \widetilde{\alpha}_{\omega, A}(\bar{t}) \cdot Q(\bar{t})]\,\delta\theta(\bar{t})\,d\bar{t}.
\end{gather}
Similarly, one obtains that $\delta \mc{B}_Q = O(\epsilon_\omega)$, so $\mc{B}_Q$ is entirely negligible. In summary, substituting this result into \Eq{eq:aux3001} gives
\begin{gather}\label{eq:dbgo}
\delta \favr{\mc{B}} \approx \delta \mc{B}_\theta \approx \frac{1}{2}\,\int_{t_1}^{t_2} (Q^\dag \cdot \alpha_{\omega, A} \cdot Q)\,\delta\theta\,dt,
\end{gather}
where we omitted corrections $O(\epsilon_\omega)$ and, for the same reason, replaced $\widetilde{\alpha}_{\omega, A}$ with $\alpha_{\omega, A}$.

%......................................
\subsubsection{Euler-Lagrange equations}

Using $\mcu{A}^{\rm (eff)}\approx A_{\text{dr}}^{(q)} + \favr{\mc{B}}$ along with \Eqs{eq:dadr} and \eq{eq:dbgo} and also requiring $\delta \mcu{A}^{\rm (eff)} = 0$, we get the following set of ELE:
\begin{align}
\delta \theta &: \quad 
\pd_t (Q^\dag \cdot \pd_{\omega} \msf{D}_{\omega, H}\cdot Q) 
= -2 Q^\dag \cdot \msf{D}_{\omega, A}\cdot Q,\label{eq:It}\\
\delta Q^\dag &: \quad \msf{D}_{\omega, H}\cdot Q = 0,
\end{align}
plus the conjugate equation for $Q^\dag$. Note that \Eq{eq:It} implies $\alpha_{\omega, A} \lesssim \alpha_{\omega, H}/(\omega \mc{T})$, which is in agreement with the original assumption \eq{eq:alphaord}.

It is also instructive to rewrite these in terms of the unit polarization vector $\msf{e}_\omega$ and the scalar amplitude $a$, assuming $Q = \msf{e}_\omega a$: 
\begin{gather}
\msf{D}_{\omega, H} \cdot \msf{e}_\omega = 0, \label{eq:Dhe}\\
\pd_t [(\msf{e}_\omega^\dag \cdot \pd_\omega \msf{D}_{\omega, H}\cdot \msf{e}_\omega)\,a^2] 
= -2(\msf{e}_\omega^\dag\cdot \msf{D}_{\omega, A}\cdot \msf{e}_\omega)\,a^2.\label{eq:ampleq}
\end{gather}
Equation \eq{eq:Dhe} can be understood as a local dispersion relation, which determines $\msf{e}_\omega(t)$ and the local frequency $\omega (t)$. Equation \eq{eq:ampleq} can be understood as an amplitude equation. In a dissipationless medium, which we \textit{define} as a medium with $\alpha_{\omega, A} = 0$, this equation becomes $\pd_t I = 0$, which is known as the action conservation theorem \cite{foot:act}. 

For example, consider a one-dimensional stationary system. In this case, \Eq{eq:Dhe} becomes
\begin{gather}\label{eq:Dh}
\msf{D}_{\omega, H} = 0,
\end{gather}
and the amplitude equation gives $a = e^{-\gamma t}\times \text{const}$, where
\begin{gather}\label{eq:gammainst}
\gamma \doteq \msf{D}_{\omega, A}/\pd_\omega \msf{D}_{\omega, H}.
\end{gather}
(This quantity is not to be confused with $\gamma$ that was introduced in \Sec{sec:co}.) One hence obtains that $q$ has a well-defined complex frequency; \ie $q(t) = \text{Re}\,(e^{-i \omega_c t}\times \text{const})$, where $\omega_c \doteq \omega - i \gamma =\omega -i \msf{D}_{\omega, A}/\pd_\omega\msf{D}_{\omega, H}$. Within the accuracy of our theory, this can be understood as a solution of the complex dispersion relation
\begin{gather}\label{eq:DO}
\msf{D}_{\omega_c} = 0,
\end{gather}
as seen from the fact that $\msf{D}_{\omega_c} = \msf{D}_{\omega - i \gamma, H} + i \msf{D}_{\omega - i \gamma, A}\approx \msf{D}_{\omega, H} + i(-\gamma \pd_\omega \msf{D}_{\omega, H} + \msf{D}_{\omega, A})$ combined with \Eqs{eq:Dh} and \eq{eq:gammainst}. Hence, $\msf{D}_{\omega_c}$ can be understood as the complex dispersion function.

%%%%%%%%%%%%%%%%%%%%%%%%%%%%%%%%%%%%%%%
\section{Waves in distributed systems}
\label{sec:cont}

%--------------------------------------
\subsection{Basic definitions}

Although the above results were derived for discrete oscillators, they can be readily extended to waves in continuous media. A continuous medium is understood as a system where the scalar product can be expressed as an integral over some continuous space $\mc{X}$ (we assume that $\mc{X}$ is Euclidean for clarity, but see \Refs{my:amc, my:wkin} for a more general case), specifically as
\begin{gather}
a \cdot b = \int_\mc{X} \vec{a}(\vec{x}) \cdot \vec{b}(\vec{x})\, d^D x.
\end{gather}
Here, $D = \text{dim}\,\mc{X}$, and $\vec{a}(\vec{x})$ and $\vec{b}(\vec{x})$ are fields of some finite dimension $D'$. For example, they can be electromagnetic fields in physical space; then $D' = 3$. (We use bold symbols to distinguish $D$-dimensional coordinates and $D'$-dimensional vectors from $N$- and $M$-dimensional coordinates and vectors that we introduced earlier, since $N$ and $M$ are infinite in the continuous limit.) Accordingly, the polarizability operator can be rewritten~as
\begin{gather}\notag
(\vec{\alpha} \circ \vec{q})(t, \vec{x}) = 
\int^t_{t_1} dt' \int_{\mc{X}} d^D x' \,
\vec{\alpha}(t, t'; \vec{x}, \vec{x}') \cdot \vec{q}(t', \vec{x}').
\end{gather}
The symmetrized kernel is introduced as $ \bar{\vec{\alpha}} (\bar{t}, \tau; \bar{\vec{x}}, \vec{\rho}) \doteq \vec{\alpha}(t, t'; \vec{x}, \vec{x}')$, where $\bar{t}$ and $\tau$ are defined as usual [\Eq{eq:ttau}], and 
\begin{gather}
\bar{\vec{x}} \doteq (\vec{x} + \vec{x}')/2, \quad \vec{\rho} \doteq \vec{x} - \vec{x}'.
\end{gather}
We also introduce the corresponding Weyl symbol (the term is used with the same reservations as earlier \cite{foot:weyl}),
\begin{gather}\notag
\vec{\alpha}_{\omega, \vec{k}}(\bar{t}, \bar{\vec{x}}) \doteq \int_0^{\infty} d\tau \int_{\mc{X}} d^D\rho\,
\bar{\vec{\alpha}} (\bar{t}, \tau; \bar{\vec{x}}, \vec{\rho})\,e^{i \omega \tau - i \vec{k} \cdot \vec{\rho}}.
\end{gather}
The modification of the remaining notation is obvious, so it will not be presented in detail.

%--------------------------------------
\subsection{Geometrical optics}

Let us consider a special case of practical interest where the medium is weakly inhomogeneous in both time and space. (Anisotropy and general dispersion, including both temporal and spatial dispersion, are implied.) Unlike in \Sec{sec:fastglobal}, where all oscillators were assumed to have the same $\omega$ at given $t$, we now allow the frequency to be spatially inhomogeneous. Specifically, we adopt
\begin{gather}
\vec{q}(t, \vec{x}) = \text{Re} [\vec{Q}(t, \vec{x})\,e^{i \theta (t, \vec{x})}]
\end{gather}
and assume that the frequency and the wave vector,
\begin{gather}\label{eq:defwk}
\omega \doteq - \pd_t \theta, \quad \vec{k} \doteq \del \theta,
\end{gather}
are slow functions of $(t, \vec{x})$. Explicitly, we require 
\begin{gather}
\max \{\epsilon_\omega, \epsilon_{\vec{k}} \} \ll 1,
\end{gather}
where $\epsilon_\omega$ is defined in \Eq{eq:epsilonom}, and $\epsilon_{\vec{k}}$ is the analogous parameter that involves spatial scales instead of temporal scales. Then, one obtains $\smash{\mcu{A}^{\rm (eff)}\approx A_{\text{dr}}^{(q)} + \favr{\mc{B}}}$, where $A_{\text{dr}}^{(q)} = \int \mcc{L}\, d^D x$, and $\mcc{L}$ is the Lagrangian density given~by
\begin{gather}
\mcc{L}(t, \vec{x}, \vec{Q}, \vec{Q}^\dag, \omega, \vec{k})
 = \frac{1}{4}\,\vec{Q}^\dag \cdot \vec{\msf{D}}_{(\omega, \vec{k}), H}(t, \vec{x}) \cdot \vec{Q}.
\end{gather}
This leads to \cite{my:amc}
\begin{multline}\notag
\delta A_{\text{dr}}^{(q)} = \int_{t_1}^{t_2} dt \int_{\mc{X}} d^D x \bigg\{(\pd_t \mc{I} + \del \cdot \vec{\mc{J}})\, \delta \theta 
\\ + \frac{1}{4}\, \delta \vec{Q}^\dag \cdot [\vec{\msf{D}}_{(\omega, \vec{k}), H}\cdot \vec{Q}]
+ \frac{1}{4}\,[\vec{Q}^\dag \cdot \vec{\msf{D}}_{(\omega, \vec{k}), H}] \cdot \delta \vec{Q}\bigg\},
\end{multline}
where $\mc{I} \doteq \pd_{\omega}\mcc{L}$ and $\vec{\mc{J}} \doteq - \pd_{\vec{k}}\mcc{L}$. The expression for $\delta \favr{\mc{B}}$ is obtained like in \Sec{sec:fastglobal} and can be written as
\begin{gather}
\delta \favr{\mc{B}} \approx \frac{1}{2} \int_{t_1}^{t_2} dt \int d^D x \,[\vec{Q}^\dag \cdot \vec{\alpha}_{(\omega, \vec{k}), A} \cdot \vec{Q}]\,\delta \theta.
\end{gather}
Hence, one obtains the following ELE:
\begin{align}
\delta \theta &: \quad 
\pd_t \mc{I} + \del \cdot \vec{\mc{J}} = -2 [\vec{Q}^\dag \cdot \vec{\msf{D}}_{(\omega,\vec{k}), A}\cdot \vec{Q}],\label{ref:act}\\
\delta \vec{Q}^\dag &: \quad \vec{\msf{D}}_{(\omega,\vec{k}), H}\cdot \vec{Q} = 0,\label{eq:Qdisp}
\end{align}
plus the conjugate equation for $\vec{Q}^\dag$. 

It is also instructive to rewrite these equations as
\begin{gather}
\vec{\msf{D}}_{(\omega,\vec{k}), H} \cdot \vec{\msf{e}}_{\omega,\vec{k}} = 0, \label{eq:Dhe2}\\
\pd_t \vec{k} = - \del \omega, \\
\pd_t \mc{I} + \del \cdot (\vec{v}_g \mc{I}) = - 2 \gamma \mc{I}.\label{eq:ampleq2}
\end{gather}
Here, $\vec{\msf{e}}_{\omega,\vec{k}}$ is the unit polarization vector. We also added an equation for $\vec{k}$ that flows from its definition [\Eq{eq:defwk}], introduced $\vec{v}_g \doteq - \pd_{\vec{k}}\mcc{L}/\pd_\omega\mcc{L}$, which represents the group velocity, and also introduced
\begin{gather}\label{eq:gammaGO}
\gamma \doteq 
\frac{\vec{\msf{e}}_{\omega,\vec{k}}^\dag \cdot \vec{\msf{D}}_{(\omega,\vec{k}), A}\cdot \vec{\msf{e}}_{\omega,\vec{k}}}
{\vec{\msf{e}}_{\omega,\vec{k}}^\dag \cdot \pd_\omega \vec{\msf{D}}_{(\omega,\vec{k}), H}\cdot \vec{\msf{e}}_{\omega,\vec{k}}},
\end{gather}
which represents the local dissipation rate. Equations \eq{eq:Dhe2}-\eq{eq:gammaGO} form a complete set of equations that describe dissipative GO waves in inhomogeneous nonstationary media. They are in agreement with results of \Ref{my:amc}, where dissipative effects were described as an addition to the variational formulation rather than as a part of it. (The variational formulation of \textit{dissipationless} wave dynamics is also discussed in \Refs{book:whitham, book:tracy}.) 

Applications of these general equations to electromagnetic waves are discussed in detail in \Ref{my:amc}. Here, we only point out that one can choose $\vec{Q}$ to be the complex amplitude of the electric field, $\vec{E}_c$; then,
\begin{gather}\label{eq:LEM}
\mcc{L} = 
\frac{1}{16\pi}\,\vec{E}_c^* \cdot \vec{\varepsilon}_{(\omega,\vec{k}), H} \cdot \vec{E}_c 
- \frac{c^2|\vec{k} \times \vec{E}_c|^2}{16\pi \omega^2},\\
\vec{\msf{D}}_{(\omega,\vec{k}), A} = \frac{\vec{\varepsilon}_{(\omega,\vec{k}), A}}{16\pi}.
\end{gather}
The second term in \Eq{eq:LEM} is understood as $|\vec{B}_c|^2/16\pi$, where we used Faraday's law $\vec{B}_c \approx (c\vec{k}/\omega) \times \vec{E}_c$ to express the complex amplitude of the magnetic field $\vec{B}_c$, and $c$ is the speed of light. Also, $\vec{\varepsilon}_{(\omega,\vec{k})}$ is the dielectric tensor in the spectral representation (or, more specifically, the Weyl symbol of the dielectric permittivity operator), and the indexes $H$ and $A$ denote its Hermitian and anti-Hermitian parts, as usual. Accordingly, \Eq{ref:act} can be understood as a restatement of Poynting's theorem \cite{my:amc}.

As a side note, an even simpler variational derivation of the results reported in this section is possible by means of the Weyl calculus combined with a technique \textit{\`a~la}~\Ref{ref:herrera86}. We leave details to future publications.

%%%%%%%%%%%%%%%%%%%%%%%%%%%%%%%%%%%%%%%
\section{Conclusions}
\label{sec:conc}

In summary, we have formulated a variational principle for a dissipative subsystem of an overall-conservative system by introducing constant Lagrange multipliers and Lagrangians nonlocal in time. We call it the variational principle for projected dynamics, or VPPD. We have also elaborated on applications of the VPPD to the special case of linear systems, particularly in the context of wave propagation in general linear media. The focus was on how the variational formulation helps in deriving reduced models, such as the quasistatic and GO models. In particular, we have proposed a variational formulation of dissipative GO in a linear medium that is inhomogeneous, nonstationary, nonisotropic, and exhibits both temporal and spatial dispersion simultaneously. The ``spectral representation'' of the dielectric tensor $\vec{\varepsilon}_{(\omega,\vec{k})}$ that enters GO equations is shown to be, strictly speaking, the Weyl symbol of the medium dielectric permittivity. This can be considered as an invariant definition of $\vec{\varepsilon}_{(\omega,\vec{k})}$, as opposed to \textit{ad~hoc} definitions that are commonly adopted in literature and have been a source of a continuing debate. 

Our work is also intended as a stepping stone for extending the variational theory of modulational stability in general wave ensembles \cite{tex:myqponder} to dissipative systems. Details will be described in a separate publication. Also, applications to electrostatic plasma oscillations are discussed in \App{app:landau}.

The authors thank D.~B\'enisti for stimulating discussions. The work was supported by the NNSA SSAA Program through DOE Research Grant No. DE-NA0002948, by the U.S. DTRA through Grant No.  HDTRA1-11-1-0037, by the U.S. DOE through Contract No. DE-AC02-09CH11466, and by the U.S. DOD NDSEG Fellowship through Contract No. 32-CFR-168a. 

\appendix

%%%%%%%%%%%%%%%%%%%%%%%%%%%%%%%%%%%%%%%
\section{Electrostatic plasma oscillations as an example}
\label{app:landau}

Here, we discuss an example of a dispersive medium, namely, electron plasma, in order to illustrate the basic concepts and notation introduced in the main text. 

%--------------------------------------
\subsection{Single-particle Lagrangian}

First, consider a single electron. Assuming a nonrelativistic quantum model, the particle Lagrangian can be expressed as $L = \int \mcc{L}\,d^3x$, where $\mcc{L}$ is the Lagrangian density given by
\begin{gather}
\mcc{L} = \frac{i\hbar}{2}\,
\big[\phi^*(\pd_t\phi) - (\pd_t\phi^*)\phi
\big] - \phi^* \,\hat{\mc{H}} \,\phi,
\end{gather}
$\phi$ is the electron wave function, and $\smash{\hat{\mc{H}}}$ is the Hamiltonian. For simplicity, we limit our consideration to electrostatic interactions, so $\smash{\hat{\mc{H}}}$ is adopted in the following form:
\begin{gather}
\hat{\mc{H}} = -\frac{\hbar^2}{2m_e}\,\del^2 + e_e \varphi(t,\vec{x}).
\end{gather}
Here, $m_e$ and $e_e$ are the electron mass and charge, respectively, and $\varphi$ is the electrostatic potential. The function $\phi$ satisfies the Schr\"odinger equation $i \hbar\, \pd_t\phi = \smash{\hat{\mc{H}}}\phi$. 

To the zeroth order in $\varphi$, the electron eigenwaves are $\phi(t, \vec{x}) = e^{i S(t, \vec{x})/\hbar}\,\phi_0$, where $\phi_0 = \text{const}$ and $S$ satisfies
\begin{gather}
\pd_t S + \frac{(\del S)^2}{2m_e} = 0.\label{eq:Sc}
\end{gather}
Below, we study the particle dynamics linearized around such eigenwaves. Consider a variable transformation
\begin{gather}
\phi(t, \vec{x}) = e^{i S(t, \vec{x})/\hbar}\, \widetilde{\phi}(t, \vec{x}),
\end{gather}
where $\smash{\widetilde{\phi}}$ is a new independent function. Then, $\mcc{L}$ becomes
\begin{gather}
\mcc{L}=\frac{i \hbar}{2}\,\big[
\widetilde{\phi}^* (\pd_t\widetilde{\phi})- (\pd_t\widetilde{\phi}^*)\widetilde{\phi}
\,\big]
- \widetilde{\phi}^*\, \hat{\mc{H}}_{\text{eff}}\, \widetilde{\phi},\\
\hat{\mc{H}}_{\rm eff} = \pd_t S + e^{-i S/\hbar}\, \hat{\mc{H}}\, e^{i S/\hbar}.
\end{gather}
As can be shown straightforwardly,
\begin{multline}
\hat{\mc{H}}_{\rm eff} = \frac{1}{2}\,[\vec{v} \cdot (-i \hbar  \del) + (-i \hbar  \del) \cdot \vec{v}]\\
-\frac{\hbar^2}{2m_e}\,\del^2 + e _e \varphi(t, \vec{x}),
\end{multline}
where \Eq{eq:Sc} was used. The vector $\vec{v}\doteq (\del S)/m_e$ is understood as the unperturbed velocity. Since the general solution of \Eq{eq:Sc} is $S = - (m \vec{u}^2/2) t + m \vec{u} \cdot \vec{x} + \text{const}$, where $\vec{u}$ is a constant parameter, one finds that $\vec{v} = \vec{u} = \text{const}$. Thus, $\vec{v}$ commutes with $\del$, and $\mcc{L}$ is simplified as follows:
\begin{multline}
\mcc{L} = 
\frac{i\hbar}{2}\,\big[
\widetilde{\phi}^* (\pd_t\widetilde{\phi}) - (\pd_t\widetilde{\phi}^*)\widetilde{\phi}
\,\big] \\- 
\widetilde{\phi}^*\left[\vec{v}\cdot (-i \hbar  \del)-\frac{\hbar^2}{2m_e}\,\del^2 + e_e \varphi \right]\widetilde{\phi}.
\end{multline}

In order to cast this Lagrangian density in the form adopted in the main text, let us introduce the notation $\smash{\widetilde{\phi}} = (1 + \smash{\widetilde{\psi}})\phi_0$, where $\phi_0$ is constant and $\smash{\widetilde{\psi}} = O(\varphi)$. Assuming $\varphi$ is small, we use the following approximation:
\begin{gather}\widetilde{\phi}^*\, e_e \varphi\, \widetilde{\phi}
\approx |\phi_0|^2 e_e \varphi\,(1+\widetilde{\psi}^*+\widetilde{\psi}),
\end{gather}
where we neglected $O(\varphi^3)$. Then, up to a complete time derivative (which does not affect ELEs), we have
\begin{multline}
\mcc{L}=\frac{i}{2}\,\big[
\psi^*(\pd_t\psi) - (\pd_t\psi^*)\psi 
\big] - \psi^*\,\hat{H}_0\,\psi 
\\-
\frac{|\phi_0|}{\sqrt{\hbar}}\,e_e \varphi(\psi^*+\psi) - |\phi_0|^2 e_e \varphi, 
\end{multline}
where $\psi \doteq \sqrt{\hbar}\,|\phi_0|\widetilde{\psi}$ and
\begin{gather}
\hat{H}_0\doteq \vec{v}\cdot (-i \del)+ \frac{\hbar}{2m_e}\, (-i \del)^2.
\end{gather}
Although the second term contains $\hbar$, it must be retained even if one is interested in the classical limit only. This is because $\hbar$ is also contained in the interaction term that is proportional to $\varphi$.

%--------------------------------------
\subsection{Plasma Lagrangian}

Now consider a plasma comprised of $N$ electrons and background ions, which we assume to be stationary. Assuming also that the plasma is nondegenerate, the self-consistent electrostatic Lagrangian density is as follows:
\begin{multline}
\mcc{L} = \frac{(\del \varphi)^2}{8\pi} + \sum_{m=1}^N \bigg\{
\frac{i}{2}\,\big[\psi_m^*(\pd_t\psi^m) - (\pd_t\psi_m^*)\psi^m\big]
\\ - \psi_m^*\,\hat{H}_{0,m}\,\psi^m
- \frac{|\phi_{m,0}|}{\sqrt{\hbar}}\, e_e \varphi (\psi_m^*+\psi^m)
\bigg\}.
\end{multline}
(The term linear in $\varphi$ is canceled by the ion contribution.) Here, $\psi^m$ is the wave function of $m$th electron and $\psi_m^*$ is its complex conjugate. This can also be expressed as
\begin{multline}\label{eq:auxa1}
\mcc{L} = \frac{(\del \varphi)^2}{8\pi} + \frac{i}{2}\,\big[
\psi^{\dag}\cdot (\pd_t\psi) - (\pd_t\psi^{\dag})\cdot \psi
\big]\\
-\psi^{\dag}\cdot \hat{H}_0\cdot \psi +\gamma^{\dag}\cdot \psi +\psi^{\dag}\cdot \gamma, 
\end{multline}
where we introduced $\psi^{\dag} \doteq (\psi_1^*, \psi_2^*, \ldots, \psi_N^*)$,
\begin{gather}
\psi \doteq \left(
\begin{array}{c}
 \psi^1 \\
 \psi^2 \\
 \vdots \\
 \psi^N \\
\end{array}
\right), \quad
\gamma \doteq -\frac{e_e \varphi}{\sqrt{\hbar}}\left(
\begin{array}{c}
 |\phi_{1,0}| \\
 |\phi_{2,0}| \\
 \vdots \\
 |\phi_{N,0}| \\
\end{array}
\right), 
\end{gather}
and $\smash{\hat{H}_0 \doteq \text{diag}\,\{\hat{H}_{0,1}, \hat{H}_{0,2}, \ldots, \hat{H}_{0,N}\}}$. The Lagrangian \eq{eq:auxa1} is of the same type as those considered in \Sec{sec:losc}, with $\varphi$ serving as $q$. (As discussed in \Sec{sec:cont}, the spatial coordinate in distributed systems serves as a continuous index.) Then, the VPPD applies as described in the main text. There is no need to repeat the formulation here, but let us show explicitly how the electron polarizability is recovered within this formalism.

%--------------------------------------
\subsection{Polarizability}

In order to eliminate $\hat{H}_0$ from $\mcc{L}$, we introduce a variable transformation $\psi \doteq U \circ \Psi$ with $U = \text{diag}\,\{U^1, U^2, \ldots U^N\}$, where $\smash{U^m\doteq \exp(-i\hat{H}_{0,m}t)}$ are the propagators of the homogeneous equations $i \pd_t\psi^m =\hat{H}_{0,m}\psi^m$. More explicitly, $U^m$ are given by
\begin{gather}
(U^m \circ \eta)(t, \vec{x}) \equiv \int U_{\vec{x}, \tilde{\vec{x}}}^{(m)}(t)\,\eta (\tilde{\vec{x}})\,d^3\tilde{x},
\end{gather}
where $\eta$ is any function, $\smash{\int \equiv \int_{-\infty}^{\infty}\int_{-\infty}^{\infty}\int_{-\infty}^{\infty}}$, and $\smash{U_{\vec{x}, \tilde{\vec{x}}}^{(m)}}$ is the corresponding Green's function; namely,
\begin{multline}
U_{\vec{x}, \tilde{\vec{x}}}^{(m)}(t) 
= \frac{1}{(2\pi)^3} \int \exp \bigg[i \vec{\kappa}\cdot (\vec{x}-\tilde{\vec{x}})\\
-i \vec{\kappa} \cdot \vec{v}_m t - \frac{i\hbar \kappa^2 t}{2m_e}\bigg]\,d^3\kappa.
\end{multline}
Then,
\begin{multline}
\mcc{L}=\frac{(\del \varphi)^2}{8\pi} + \frac{i}{2}\,\big[
\Psi^{\dag}\cdot (\pd_t\Psi) - (\pd_t\Psi^{\dag}) \cdot \Psi
\big]\\
+\Gamma^{\dag}\cdot \Psi +\Psi^{\dag}\cdot \Gamma,
\end{multline}
where $\Gamma$ is a vector with the following components:
\begin{multline}
\Gamma^m\doteq U^{m\dag}  \circ \gamma 
= -\frac{|\phi_{0,m}|e_e}{\sqrt{\hbar}}\,(U^{m \dag} \circ \varphi)\\
=-\frac{|\phi_{0,m}|e_e}{\sqrt{\hbar}}\int U_{\vec{x}', \vec{x}}^{(m)*}(t)\,\varphi (t, \vec{x}')\,d^3x'.
\end{multline}
This corresponds to the following matrix $V$ [\Eq{eq:Vdef}]:
\begin{gather}
V_{\vec{x}, \vec{x}'}^{(m)}(t)=-\sqrt{\frac{2}{\hbar}}\,|\phi_{0,m}| e_e U_{\vec{x}', \vec{x}}^{(m)*}(t).
\end{gather}
Hence, the polarizability kernel [\Eq{eq:akern}] is as follows:
\begin{align}
\alpha & (t, \vec{x}, t', \vec{x}') \notag\\
& =-\sum_{m=1}^N \text{Im} \int [V^{(m)\dag}]_{\vec{x}, \tilde{\vec{x}}}(t)\cdot V_{\tilde{\vec{x}}, \vec{x}'}^{(m)}(t')\,d^3\tilde{x} \notag\\
&=-\sum_{m=1}^N \text{Im} \int V_{\tilde{\vec{x}}, \vec{x}}^{(m)*}(t) \cdot V_{\tilde{\vec{x}}, \vec{x}'}^{(m)}(t')\,d^3\tilde{x} \notag\\
&=-\frac{2e_e^2}{\hbar} \sum_{m=1}^N  |\phi_{0,m}|^2\,
\text{Im}\,\int U_{\vec{x}, \tilde{\vec{x}}}^{(m)}(t)\,\cdot U_{\vec{x}', \tilde{\vec{x}}}^{(m)*}(t')\,d^3\tilde{x}\notag\\
&=-\frac{2n_0e_e^2}{\hbar}\,\text{Im}\left\langle \msf{J}_m\right\rangle_{\vec{v}}.
\end{align}
Here, we introduced the unperturbed density $n_0\doteq \smash{\sum_{m=1}^N |\phi_{0,m}|^2}$, the ensemble averaging
\begin{gather}
\favr{(\ldots)_m}_{\vec{v}} \doteq \frac{1}{n_0} \sum_{m=1}^N |\phi_{0,m}|^2(\ldots)_m,
\end{gather}
and also the following quantity (where the index $m$ is omitted for brevity):
\begin{multline}\notag
\msf{J} 
\doteq \int U_{\vec{x}, \tilde{\vec{x}}}(t)\cdot U_{\vec{x}', \tilde{\vec{x}}}^*(t')\,d^3\tilde{x} 
\\ 
= \int e^{i \vec{\kappa}\cdot (\vec{x}-\vec{x}') -i \vec{\kappa} \cdot \vec{v}(t-t')- i \hbar  \kappa^2 (t-t')/(2m_e)}\,
\frac{d^3\kappa}{(2\pi)^3}.
\end{multline}
The corresponding symmetrized kernel (\Sec{sec:symm}) is
\begin{gather}
\bar{\alpha}\left(\bar{t}, \bar{\vec{x}}, \tau, \vec{\rho}\right)
= n_0 \favr{\bar{\alpha}^{(\vec{v})}(\tau, \vec{\rho})}_{\vec{v}},
\end{gather}
where $\bar{\alpha}_{\vec{v}}$ is the per-particle polarizability given by
\begin{gather}\notag
\bar{\alpha}^{(\vec{v})}(\tau, \vec{\rho})
= -\frac{2e_e^2}{\hbar}
\int \sin \left[
\vec{\kappa}\cdot \vec{\rho}-  \vec{\kappa} \cdot \vec{v} \tau -\frac{\hbar \kappa^2 \tau}{2m_e}
\right]\frac{d^3\kappa}{(2\pi)^3}.
\end{gather}
Its spectral representation is as follows:
\begin{multline}
\alpha_{\omega, \vec{k}}^{(\vec{v})} = \int_0^{\infty}d\tau\,  e^{i \omega  \tau} \int d^3\rho\,
e^{-i \vec{k}\cdot \vec{\rho}}\,
\bar{\alpha}^{(\vec{v})}(\tau, \vec{\rho})\\
=\frac{2e_e^2}{\hbar} \int_0^{\infty} e^{i (\omega - \vec{k}\cdot \vec{v}) \tau}\sin \left(\frac{\hbar k^2 \tau}{2m_e}\right)d\tau,
\end{multline}
where $\text{Im}\, \omega > 0$ is assumed. Hence, one obtains
\begin{gather}
\alpha_{\omega, \vec{k}}^{(\vec{v})} = -\frac{e_e^2}{m_e}\frac{k^2}{(\omega -\vec{k}\cdot \vec{v})^2-\hbar^2k^4/\left(4m_e^2\right)}. \end{gather}

This result is in agreement with the one that was presented in \Ref{tex:myqponder} for adiabatic interactions. The additional factor $k^2$ is due to the fact that $\smash{\alpha_{\omega, \vec{k}}^{(\vec{v})}}$ defines the particle response with respect to the potential $\varphi$ rather than with respect to the electric field $\vec{E} = -\del \varphi$ as usual. The corresponding susceptibility (in the electrostatic approximation assumed here)~is 
\begin{align}
\chi_{\omega, \vec{k}}
& = \frac{4\pi n_0}{k^2} \favr{\alpha_{\omega, \vec{k}}^{(\vec{v})}}_\vec{v} \notag \\
& = - \omega_p^2 \int \frac{F_0(\vec{v})}{(\omega-\vec{k}\cdot \vec{v})^2-\hbar^2 k^4/(4m_e^2)}\,d^3v \notag\\
& = - \frac{\omega_p^2}{k^2}\int_{-\infty}^\infty \frac{f_0(v_\lVert)}{(v_\lVert - \omega/k)^2 - \hbar^2 k^2/(4m_e^2)} \, dv_\lVert.\notag
\end{align}
Here, $\omega_p \doteq (4\pi n_0 e_e^2/m_e)^{1/2}$, $F_0$ is the velocity distribution, $f_0(v_\lVert) \doteq \smash{\int F_0(\vec{v}_\perp, v_\lVert)\,d^2v_\perp}$, $\vec{v}_\perp$ is the component of $\vec{v}$ transverse to $\vec{k}$, and $v_\lVert \doteq \vec{k} \cdot \vec{v}/k$. 

The standard classical result \cite{book:stix} is obtained by taking $\hbar \to 0$ and integrating by parts. In particular, for real $\omega$, the susceptibility of classical electron plasma is
\begin{align}
\chi_{\omega, \vec{k}}
& = - \frac{\omega_p^2}{k^2}\int_{-\infty}^\infty \frac{f_0(v)}{[v-(\omega +i 0)/k]^2} \, dv\notag\\
& = - \frac{\omega_p^2}{k^2}\int_{-\infty}^\infty \frac{f_0'(v)}{v-(\omega +i 0)/k} \, dv\notag\\
& = - \frac{\omega_p^2}{k^2}\,\msf{PV}\int_{-\infty}^\infty \frac{f_0'(v)}{v-\omega/k} \, dv
-i \pi\, \frac{\omega_p^2}{k^2}\,f_0'\left(\frac{\omega}{k}\right)\text{sgn}\,k,\notag
\end{align}
where $\msf{PV}$ denotes the Cauchy principal value. The last term determines the anti-Hermitian (imaginary) part of $\chi_{\omega, \vec{k}}$, which is responsible for Landau damping \cite{book:stix}. In this model, when $f'_0(\omega/k) = 0$, the plasma response is adiabatic and the usual variational formulation applies which only involves local Lagrangians \cite{my:itervar}. If $f'_0(\omega/k)$ is nonzero, a nonlocal Lagrangian must be used as described in the main text of the present paper. Also note that the quasistatic limit discussed in \Sec{sec:qs} corresponds to vanishingly small $\omega/k$. Then, $\chi_{\omega, \vec{k}} = \smash{(k\lambda_D)^{-2}}$, where $\smash{\lambda_D^{-2}} \doteq -\omega_p^2\, \smash{\msf{PV}\int [f_0'(v)/v]\,dv}$. In this regime, the plasma response manifests as Debye shielding \cite{book:stix}.

%%%%%%%%%%%%%%%%%%%%%%%%%%%%%%%%%%%%%%%
\section{Compact form of the phase-space action}
\label{app:psp}

As a side note, equations of \Sec{sec:sym} can be expressed in a more compact form in terms of the phase space coordinate $z$ and the Poincar\'e two-form~$\omega_{\alpha\beta}$,
\begin{gather}
z \doteq \left(
\begin{array}{c}
\xi\\[3pt]
\eta
\end{array}
\right),
\quad
\omega_{\alpha\beta} \doteq 
\left(
\begin{array}{c @{\quad} c}
0 & - \mathbb{I}\\[3pt]
 \mathbb{I} & 0
\end{array}
\right),
\end{gather}
where $\mathbb{I}$ is a $M \times M$ unit matrix. In particular, one can cast \Eq{eq:aux33} as
\begin{gather}
\mcu{A}[q, z] = \int^{t_2}_{t_1} \Big[\Lambda^{(q)} 
+ \frac{1}{2}\,\omega_{\alpha\beta} z^\alpha \dot{z}^\beta - H^{(z)}\Big]\,dt
\end{gather}
(Greek indexes span from 1 to $2M$) and \Eq{eq:aux34} as
\begin{gather}
\delta \mcu{A}[q, z] = \Big[p \cdot \delta q + \frac{1}{2}\,\omega_{\alpha\beta}z^\alpha\,\delta z^\beta\Big]\,\Big|_{t_1}^{t_2} + \delta \mcu{F}^{(q)} + \delta \mcu{F}^{(z)}, \notag\\
\delta \mcu{F}^{(z)} \doteq \int^{t_2}_{t_1}(\omega_{\alpha\beta}\dot{z}^\beta - \pd_\alpha H^{(z)})\,\delta z^\alpha\,dt,
\end{gather}
where $\pd_\alpha \doteq \pd/\pd z^\alpha$ and $H^{(z)} (t, z) \doteq H^{(\xi)}(t, \xi, \eta)$. Accordingly, Hamilton's equations \eq{eq:he} are cast as follows:
\begin{gather}\label{eq:hez}
\omega_{\alpha\beta}\dot{z}^\beta = \pd_\alpha H^{(z)}.
\end{gather}
Also, \Eq{eq:aeffsr} can be represented as follows:
\begin{gather}\label{eq:auxx}
\mcu{A}^{\rm (eff)}[q]  = \mcu{A}[q, \hat{z}] 
+ \frac{1}{2}\,\omega_{\alpha\beta} \hat{z}^\alpha(t_2) z^\beta_T,
\end{gather}
where 
\begin{gather}
z_T \doteq \left(
\begin{array}{c}
\xi_T\\[3pt]
\eta_T
\end{array}
\right).
\end{gather}
On physical trajectories one has $\hat{z}^\alpha(t_2) = z^\alpha_T$. Since $\omega_{\alpha\beta}$ is antisymmetric, this implies that the second term on the right side of \Eq{eq:auxx} vanishes on such trajectories.

%%%%%%%%%%%%%%%%%%%%%%%%%%%%%%%%%%%%%%%
\section{Coupling to adiabatic oscillators}
\label{app:osc}

Here, we discuss an instructive special case, namely, the situation when the Hamiltonian $\smash{H^{(\xi)}}$ in \Eq{eq:apq} does not depend on $\xi$ explicitly. In this case, $\eta$ is conserved and can be treated as a constant parameter, which we denote $J$. Hence, one can write
\begin{gather}\label{eq:aeffl}
A^{\rm (eff)}[q] = \int_{t_1}^{t_2} [\Lambda^{(q)}(t, q, \dot{q}) - H^{(\xi)}(t, q, \dot{q}; J)]\,dt.
\end{gather}
An alternative derivation of the same result, also known as the Routh reduction can be found, \eg in \Ref{my:acti}.

In particular, \Eq{eq:aeffl} can be applied \cite{foot:appl} when $\mcu{S}^{(\xi)}$ is an \textit{adiabatic oscillator} in which $(\xi, \eta)$ are the angle-action variables. (Then, $J$ is called an adiabatic invariant \cite{book:landau1}.)  In the special case when $\mcu{S}^{(\xi)}$ is a \textit{linear} oscillator, one can further write $H^{(\xi)} = J \cdot \Omega$, where $\Omega$ is the corresponding canonical frequency (\ie such that $\dot{\xi} = \Omega$), which is independent of $J$ but may depend on $(t, q, \dot{q})$. Then, 
\begin{gather}\label{eq:leff}
A^{\rm (eff)}[q] = \int_{t_1}^{t_2} [\Lambda^{(q)}(t, q, \dot{q}) - J \cdot \Omega(t, q, \dot{q})]\,dt.
\end{gather}

For example, one can use this to describe a charged particle in a weakly inhomogeneous field $B$. In this case, $\Omega$ is the gyrofrequency, and $J$ is proportional to the particle magnetic moment $\mu$, such that $J \cdot \Omega = \mu B$ \cite{my:mneg}. Then, the last term in \Eq{eq:leff} is nothing but the effective potential responsible for the diamagnetic force. 

%%%%%%%%%%%%%%%%%%%%%%%%%%%%%%%%%%%%%%%
\section{Parametric and quantum interactions}
\label{app:q}

%--------------------------------------
\subsection{Basic equations}

In addition to the linear coupling discussed in \Sec{sec:co}, it is instructive to consider parametric interactions, particularly because they subsume quantum interactions and thus can be understood as most general. Like in \Sec{sec:losc}, we assume that $H_0$ is eliminated by variable transformation $\psi \mapsto \Psi$. For the interaction Hamiltonian, we adopt
\begin{gather}\label{eq:hint2}
H^{\rm (int)} = \Psi^\dag \cdot h(t, q) \cdot \Psi,
\end{gather}
where $h$ is some Hermitian operator. (We assumed that $H^{\rm (int)}$ is independent of derivatives of $q$ for brevity, but the corresponding generalization is straightforward.) The total Lagrangian in this case can be written as
\begin{gather}\label{eq:aux103}
L = \Lambda^{(q)} (t, q, \dot{q})
+ \frac{i}{2}\,(\Psi^\dag \cdot \dot{\Psi} - \dot{\Psi}^\dag \cdot \Psi) - \Psi^\dag \cdot h(t, q) \cdot \Psi,
\end{gather}
and the corresponding ELE are
\begin{align}
\delta q: \quad  & \dot{p} = \pd_q \Lambda^{(q)} - \Psi^\dag\cdot  \pd_q h \cdot \Psi,\label{eq:qleffq}\\
\delta \Psi^\dag : \quad & i\dot{\Psi} = h \cdot \Psi.\label{eq:psiQ2}
\end{align}
Notably, \Eq{eq:psiQ2} conserves $\Psi^\dag \cdot \Psi$ even in the presence of coupling, in contrast with \Eq{eq:psiQ}.

For projected dynamics, the effective action $\mcu{A}^{\rm (eff)}$ is introduced just like in \Sec{sec:pdaeff}. When the solution for $\Psi$ and $\Psi^\dag$ are substituted into \Eq{eq:aux103}, the second term and the third term cancel each other due to \Eq{eq:psiQ2}, so one ends up with
\begin{gather}
\mcu{A}^{\rm (eff)}[q] = \int^{t_2}_{t_1} \Lambda^{(q)} (t, q, \dot{q})\,dt  + \mc{B}, \\
\mc{B} \doteq \frac{i}{2}\,\big[\nu^\dag \cdot \hat{\Psi}(t_2) - \hat{\Psi}^\dag(t_2) \cdot \nu\big].
\end{gather}

%--------------------------------------
\subsection{Derivation of the ELE for $\boldsymbol{q}$}

Let us show how a correct ELE for $q$ is obtained from here by a straightforward calculation. First,
\begin{gather}
\delta\mcu{A}^{\rm (eff)}[q] = \int_{t_1}^{t_2} [\pd_q \Lambda^{(q)} - \dot{p}]\,\delta q\,dt + \delta\mc{B},\label{eq:daeffq}\\
\delta \mc{B} = \frac{i}{2}\,\big[\nu^\dag \cdot \delta \hat{\Psi}(t_2) - \delta \hat{\Psi}^\dag(t_2) \cdot \nu\big].
\end{gather}
In order to calculate $\delta \hat{\Psi}(t_2)$, let us use the fact that $\hat{\Psi}(t_2) = U_h(t_2, t_1) \cdot \Psi_0$, where $U_h$ is a unitary propagator given by the following time-ordered exponential:
\begin{gather}
U_h(t_2, t_1) = \msf{T}\,\exp\left[-i \int^{t_2}_{t_1} h(t', q(t'))\,dt' \right].
\end{gather}
If the variation $\delta q$ is localized to an infinitesimal time interval $\Delta t$ following some time $\tau$, one can readily write
\begin{gather}
\delta \hat{\Psi}(t_2) = U_h(t_2, \tau + \Delta t) \cdot \delta U_h(\tau + \Delta t, \tau) \cdot U_h(\tau, t_1) \cdot \Psi_0,\notag\\
\delta U_h(\tau + \Delta t, \tau) \approx -i\pd_q h(\tau, q(\tau)) \,\delta q(\tau)\,\Delta t, \notag
\end{gather}
where one can further substitute $U_h(\tau, t_1) \cdot \Psi_0 = \hat{\Psi}(\tau)$. Then, for a general variation $\delta q$, one gets
\begin{gather}
\delta \hat{\Psi}(t_2) = -i\int_{t_1}^{t_2} \big[U_h(t_2, \tau) \cdot \pd_q h(\tau, q(\tau)) \cdot \hat{\Psi}(\tau)\big]\,\delta q(\tau)\,d\tau.\notag
\end{gather}
For $\delta \mc{B}/\delta q$ at any given time $t$, this gives
\begin{gather}
\frac{\delta \mc{B}}{\delta q} = \text{Re}\,\big[\nu^\dag \cdot U_h(t_2, t) \cdot \pd_q h(t, q(t)) \cdot \hat{\Psi}(t)\big].
\end{gather}
Then, after substituting \Eq{eq:mupsi}, one gets
\begin{multline}
\frac{\delta \mc{B}}{\delta q} = - \text{Re}\,\big[\hat{\Psi}^\dag(t_2) \cdot U_h(t_2, t) \cdot \pd_q h(t, q(t)) \cdot \hat{\Psi}(t)\big]
\\ = - \text{Re}\,\big[\hat{\Psi}^\dag(t) \cdot \pd_q h(t, q(t)) \cdot \hat{\Psi}(t)\big].\label{eq:aux112}
\end{multline}
Here, we also used $\hat{\Psi}(t_2) = U_h(t_2, t) \cdot \Psi(t)$ and the fact that $[U_h(t_2, t)]^\dag \cdot U_h(t_2, t)$ is a unit operator, because $U_h$ is unitary. Since $\pd_q h(t, q(t))$ is Hermitian, one can omit ``Re'', so \Eq{eq:aux112} becomes
\begin{gather}\label{eq:aux110}
\frac{\delta \mc{B}}{\delta q} = - \hat{\Psi}^\dag(t) \cdot \pd_q h(t, q(t)) \cdot \hat{\Psi}(t).
\end{gather}
After substituting this into \Eq{eq:daeffq} and requiring $\delta \mcu{A}^{\rm (eff)}[q] = 0$, one obtains
\begin{gather}
\delta q: \quad  \dot{p} = \pd_q \Lambda^{(q)} - \hat{\Psi}^\dag\cdot \pd_q h \cdot \hat{\Psi}.
\end{gather}
This is equivalent to \Eq{eq:qleffq}, as expected.

%--------------------------------------
\subsection{Approximate Lagrangian of weak coupling}

Suppose that the coordinate $q$ is chosen such that $q = O(a)$, where $a$ is a small parameter. Assuming that the interaction is weak, a typical interaction Hamiltonian is then linear in $q$. Like $H_0$, any zeroth-order term can be removed by a variable transformation, so we assume $h(t, q) = w \cdot q$, where $w = O(1)$ may depend only on $t$, if at all. We now seek to derive an approximation of $L$ accurate up to $O(a^2)$; \ie terms $o(a^2)$ will be neglected. 

To construct such an approximation, we adopt an asymptotic representation of $\Psi$ as a power series in $a$; \ie $\Psi = \sum_n \Psi_n$, where $\Psi_n = O(a^n)$. According to \Eq{eq:psiQ2}, $\dot{\Psi}_n = o(a^n)$, so truncating this series as $\Psi = \Psi_0 + \Psi_1$ is sufficient. Since $\Psi_0$ is constant, $\mcu{S}^{(\xi)}$ is then fully described by $\Psi_1$, which serves as the new independent variable instead of the original $\Psi$. Hence, after omitting complete time derivatives (which do not affect ELE), \Eq{eq:aux103} becomes similar to \Eq{eq:LGamma}, namely,
\begin{multline}
L = \Lambda^{(q)}_{0} (t, q, \dot{q})
+ \frac{i}{2}\,(\Psi_1^\dag \cdot \dot{\Psi}_1 - \dot{\Psi}^\dag_1 \cdot \Psi_1) \\
+ \Gamma^\dag(t, q) \cdot \Psi_1 + \Psi_1^\dag \cdot \Gamma(t, q),
\end{multline}
where the new $\Lambda^{(q)}_{0}$ and $\Gamma$ are given by
\begin{gather}
\Lambda^{(q)}_{0}(t, q, \dot{q}) \doteq \Lambda^{(q)}(t, q, \dot{q}) - \Psi_0^\dag \cdot h(t, q) \cdot \Psi_0, \\
\Gamma(t, q) \doteq - h(t, q) \cdot \Psi_0.
\end{gather}
(One may also recognize this as the leading-order Born approximation of the original problem.) This result shows that theory of linear dispersion for interaction Hamiltonians of the form \eq{eq:hint2} can be constructed identically to that for interaction Hamiltonians of the form \eq{eq:hint}. In particular, this means that general predictions of linear dispersion theory are independent of whether oscillators comprising a medium are classical or quantum.

%%%%%%%%%%%%%%%%%%%%%%%%%%%%%%%%%%%%%%%
\section{Properties of $\boldsymbol{\widetilde{\alpha}_\omega}$ and $\boldsymbol{\alpha_\omega}$}
\label{app:alphas}

Here, we discuss some properties of the functions $\widetilde{\alpha}_\omega$ and $\alpha_\omega$ that are introduced in \Sec{sec:weyl}. 

%--------------------------------------
\subsubsection{Function $\tilde{\alpha}_\omega$}

First, \Eq{eq:aproph} is proved as follows:
\begin{align}
& \int_{-\tau_m}^{\tau_m} d\tau\,\bar{\alpha}(\bar{t}, \tau)\,\text{sgn}(\tau)\,e^{i \omega \tau}\notag\\
& = \int_0^{\tau_m} d\tau\, \bar{\alpha}(\bar{t}, \tau)\,e^{i \omega \tau} 
    - \int_{-\tau_m}^0 d\tau\, \bar{\alpha}(\bar{t}, \tau)\,e^{i \omega \tau}\notag\\
& = \int_0^{\tau_m} d\tau\, \bar{\alpha}(\bar{t}, \tau)\,\text{sgn} (\tau)\,e^{i \omega \tau}
    + \int_{\tau_m}^0 d\tau\, \bar{\alpha}(\bar{t}, -\tau)\,e^{-i \omega \tau} \notag\\
& = \int_0^{\tau_m} d\tau\,\bar{\alpha}(\bar{t}, \tau)\,e^{i \omega \tau}
    - \int_0^{\tau_m} d\tau\,\bar{\alpha}(\bar{t}, -\tau)\,e^{-i \omega \tau}\notag \\
& = \int_0^{\tau_m} d\tau\,\bar{\alpha}(\bar{t}, \tau)\,e^{i \omega \tau}
    + \int_0^{\tau_m} d\tau\,\bar{\alpha}^{\rm T}(\bar{t}, \tau)\,e^{-i \omega \tau}\notag \\
& = \widetilde{\alpha}_{\omega}(\bar{t}) + \widetilde{\alpha}_{\omega}^{\dag}(\bar{t})
 = 2\widetilde{\alpha}_{\omega, H}(\bar{t}),
\end{align}
where we used that $\bar{\alpha}$ is real by definition. Equation \eq{eq:apropa} is proved similarly.

Second, let us explicitly derive an approximation for $\widetilde{\alpha}_\omega$ in the limit \eq{eq:vro}. Using \Eq{eq:avU} for $\bar{\alpha}$ together with \Eq{eq:mcUasym} for $\mc{U}$, we obtain
\begin{multline}
\bar{\alpha}(\bar{t}, \tau)\,e^{i\omega \tau} = \frac{i}{2}\,\big[
v^\dag(\bar{t} + \tau/2) \cdot e^{i(\omega - H_0)\tau} \cdot v(\bar{t} - \tau/2)
\\- v^{\rm T}(\bar{t} + \tau/2) \cdot e^{i(\omega + H_0)\tau} \cdot v^*(\bar{t} - \tau/2)
\big].\notag
\end{multline}
Then, using $v(\bar{t} \pm \tau/2) \approx v(\bar{t})$, one gets
\begin{multline}
\widetilde{\alpha}_{\omega, H} =\frac{1}{2}\,v^\dag \cdot 
\bigg\{
\frac{1- \cos[(H_0 - \omega)\tau_m]}{H_0 - \omega} 
\\
+
\frac{1- \cos[(H_0 + \omega)\tau_m]}{H_0 + \omega} 
\bigg\} \cdot v,\notag
\end{multline}
\begin{gather}
\widetilde{\alpha}_{\omega, A} = \frac{1}{2}\, v^\dag \cdot \left\{
\frac{\sin[(H_0 - \omega)\tau_m]}{H_0 - \omega} 
-
\frac{\sin[(H_0 + \omega)\tau_m]}{H_0 + \omega} 
\right\} \cdot v,\notag
\end{gather}
or, in the continuous-spectrum limit (\Sec{sec:pm}),
\begin{gather}
\widetilde{\alpha}_{\omega, H} =
\int
\frac{1 - \cos[(\Omega - \omega)\tau_m]}{2(\Omega - \omega)} \,[f(\bar{t}, \Omega) - f(\bar{t}, -\Omega)]\,d\Omega,\label{eq:ath}\\
\widetilde{\alpha}_{\omega, A} = \int
\frac{\sin[(\Omega - \omega)\tau_m]}{2(\Omega - \omega)} \,[f(\bar{t}, \Omega) - f(\bar{t}, -\Omega)]\,d\Omega.\label{eq:ata}
\end{gather}
Notably, the anti-Hermitian part of $\widetilde{\alpha}_\omega$ vanishes in the zero-$\omega$ limit; \ie $\widetilde{\alpha}_{0, A} = 0$, so $\widetilde{\alpha}_0 \approx \widetilde{\alpha}_{0, H}$, and
\begin{multline}\label{eq:zerow}
\widetilde{\alpha}_{0, H} \approx v^\dag \cdot H_0^{-1}\cdot\big [1- \cos(H_0\tau_m)\big] \cdot v
\\ \to \int \Omega^{-1}[1 - \cos(\Omega \tau_m)]\,f(\bar{t}, \Omega)\,d\Omega. 
\end{multline}

%--------------------------------------
\subsubsection{Function $\alpha_\omega$}

Asymptotic expressions for $\alpha_\omega$ under the condition \eq{eq:vro} can be obtained by taking the large-$\tau_m$ limit of \Eqs{eq:ath} and \eq{eq:ata}. The oscillating terms vanish for $\Omega \ne \omega$. For $\Omega = \omega$, one has $1 - \cos[(\Omega - \omega)\tau_m] \equiv 0$, so, instead of averaging the corresponding cosine to zero, we entirely exclude the point $\Omega = \omega$ from the integration domain. Also, $\sin[(\Omega - \omega)\tau_m]/(\Omega - \omega) \to \pi \delta(\Omega - \omega)$. Hence, one gets
\begin{gather}
\alpha_{\omega, H} = \msf{PV} \int \frac{f(\bar{t}, \Omega) - f(\bar{t}, -\Omega)}{2(\Omega - \omega)}\,d\Omega,\label{eq:ahaa}\\
\alpha_{\omega, A} = \frac{\pi}{2}\,[f(\bar{t}, \omega) - f(\bar{t}, -\omega)],
\end{gather}
where $\msf{PV}$ denotes the Cauchy principal value. Equation \eq{eq:ahaa} can be understood also as $\alpha_{\omega, H} = -\msf{H} \alpha_{\omega, A}$, where $\msf{H}$ is the Hilbert transform. Using that $\msf{H}^2 = - 1$, we then obtain $\alpha_{\omega, A} = \msf{H} \alpha_{\omega, H}$. Thus, in summary,
\begin{gather}
\alpha_{\omega, H} = \frac{1}{\pi}\,\msf{PV} \int \frac{\alpha_{\omega, A}}{\Omega - \omega}\,d\Omega,\\
\alpha_{\omega, A} = - \frac{1}{\pi}\,\msf{PV} \int \frac{\alpha_{\omega, H}}{\Omega - \omega}\,d\Omega,
\end{gather}
which can be recognized as the Kramers-Kronig relations.

\begin{widetext}
%%%%%%%%%%%%%%%%%%%%%%%%%%%%%%%%%%%%%%%
\section{Calculation of $\boldsymbol{\delta\favr{\mc{B}}}$}
\label{app:dB}

To calculate $\delta\favr{\mc{B}}$ used in \Sec{sec:fastglobal}, we proceed as follows. First, using \Eqs{eq:auxB} and \eq{eq:nupm}, we get
\begin{gather}
\delta \mc{B}_+ 
= \text{Re} \left\{-\frac{i}{8} \int_{t_1}^{t_2}dt \int_{t_1}^{t_2}dt'\,e^{-i \theta (t)} 
  Q^{\dag}(t)\cdot V^{\dag}(t)\cdot V(t') \cdot \delta [Q(t')e^{i \theta (t')}]
  \right\}.
\end{gather}
Using the notation $E(\bar{t}, \tau) \doteq e^{i \theta (t') - i \theta (t)}$, we further rewrite this as follows:
\begin{align}
\delta \mc{B}_+
& = \frac{1}{8}\, \text{Re}\left\{
    \int_{t_1}^{t_2} dt \int_{t_1}^{t_2} dt'\,
    E(\bar{t}, \tau)\,Q^{\dag}(t)\cdot V^{\dag}(t)\cdot V(t') \cdot [Q(t')\,\delta \theta (t') - i \delta Q(t')]
    \right\} 
    \notag\\
& = \frac{1}{16} \bigg\{\int_{t_1}^{t_2} dt \int_{t_1}^{t_2} dt'\,
    E(\bar{t}, \tau)\,Q^{\dag}(t) \cdot V^{\dag}(t) \cdot V(t') \cdot Q(t')\,\delta \theta (t') 
 + \int_{t_1}^{t_2} dt \int_{t_1}^{t_2} dt'\,
    [E(\bar{t}, \tau)\,Q^{\dag}(t)\cdot V^{\dag}(t)\cdot V(t')\cdot Q(t')]^{\dag}\,\delta \theta (t')
    \notag \\ &  \mbox{} \quad  
    - i\int_{t_1}^{t_2} dt \int_{t_1}^{t_2} dt'\,
    E(\bar{t}, \tau)\,Q^{\dag}(t)\cdot V^{\dag}(t)\cdot V(t')\cdot \delta Q(t')
    + i\int_{t_1}^{t_2} dt \int_{t_1}^{t_2} dt'\,
    [E(\bar{t}, \tau)\,Q^{\dag}(t)\cdot V^{\dag}(t)\cdot V(t')\cdot \delta Q(t')]^{\dag}
    \bigg\}
    \notag \\
& = \frac{1}{16}\bigg\{
    \int_{t_1}^{t_2} dt \int_{t_1}^{t_2} dt'\,
    E(\bar{t}, \tau)\,Q^{\dag}(t)\cdot V^{\dag}(t)\cdot V(t')\cdot Q(t')\,\delta \theta (t')
    + \int_{t_1}^{t_2} dt \int_{t_1}^{t_2} dt'\,
    E(\bar{t}, -\tau)\,Q^{\dag}(t') \cdot V^{\dag}(t') \cdot V(t)\cdot Q(t)\,\delta \theta (t')
    \notag \\ &  \mbox{} \quad  
    - i\int_{t_1}^{t_2} dt \int_{t_1}^{t_2} dt'\,
    E(\bar{t}, \tau)\,Q^{\dag}(t) \cdot V^{\dag}(t)\cdot V(t')\cdot \delta Q(t')
    + i\int_{t_1}^{t_2} dt \int_{t_1}^{t_2} dt'\,
    E(\bar{t}, -\tau)\,\delta Q^{\dag}(t')\cdot V^{\dag}(t')\cdot V(t)\cdot Q(t)
    \bigg\}
    \notag \\
& = \frac{1}{16}\bigg\{
     \int_{t_1}^{t_2} dt \int_{t_1}^{t_2} dt'\,
     E(\bar{t}, \tau)\,Q^{\dag}(t)\cdot V^{\dag}(t)\cdot V(t') \cdot Q(t')\,\delta \theta (t')
     + \int_{t_1}^{t_2} dt' \int_{t_1}^{t_2} dt\,
     E(\bar{t}, \tau)\,Q^{\dag}(t)\cdot V^{\dag}(t)\cdot V(t')\cdot Q(t')\,\delta \theta (t)
     \notag \\ &  \mbox{} \quad  
     - i\int_{t_1}^{t_2} dt \int_{t_1}^{t_2} dt'\,
     E(\bar{t}, \tau)\,Q^{\dag}(t)\cdot V^{\dag}(t)\cdot V(t')\cdot \delta Q(t')
     + i\int_{t_1}^{t_2}dt \int_{t_1}^{t_2} dt'\,
     E(\bar{t}, \tau)\,\delta Q^{\dag}(t)\cdot V^{\dag}(t)\cdot V(t')\cdot Q(t')
    \bigg\}
    \notag \\
& = \frac{1}{8}\bigg\{
    \int_{t_1}^{t_2} dt \int_{t_1}^{t_2} dt'\,
    E(\bar{t}, \tau)\,Q^{\dag}(t) \cdot V^{\dag}(t)\cdot V(t')\cdot Q(t')\,\frac{\delta \theta (t')+\delta \theta(t)}{2}
    \notag \\ &  \mbox{} \quad  
    -\frac{i}{2}\int_{t_1}^{t_2} dt \int_{t_1}^{t_2} dt'\,
    E(\bar{t}, \tau)\,Q^{\dag}(t) \cdot V^{\dag}(t) \cdot V(t')\cdot \delta Q(t')
    + \frac{i}{2}\int_{t_1}^{t_2} dt \int_{t_1}^{t_2} dt'\, E(\bar{t}, \tau)\,\delta Q^{\dag}(t)\cdot V^{\dag}(t)\cdot V(t')\cdot Q(t')
    \bigg\}
    \notag \\
& = \frac{1}{8}\,\text{Tr} \int_{t_1}^{t_2} dt \int_{t_1}^{t_2} dt'\,
     E(\bar{t}, \tau) \left\{W(\bar{t}, \tau)\,
     \frac{\delta \theta (t')+\delta \theta (t)}{2}
     + \frac{i}{2} \left[Q(t')\cdot \delta Q^\dag (t)-\delta Q(t')\cdot Q^\dag (t)\right]
     \right\}\cdot V^{\dag}(t)\cdot V(t').\notag
\end{align}
Next, we notice that $\delta \mc{B}_-[Q, \theta] = \delta \mc{B}_+[Q^*, -\theta] = (\delta \mc{B}_+[Q^*, -\theta])^*$. Hence,
\begin{multline}
\delta \mc{B}_- 
= \frac{1}{8}\,\text{Tr}\left[\int_{t_1}^{t_2} dt \int_{t_1}^{t_2} dt'\,
  E(\bar{t}, -\tau) \left\{
  W^*(\bar{t}, \tau)\frac{-\delta \theta (t') - \delta \theta (t)}{2} 
  + \frac{i}{2} \left[Q^*(t')\cdot \delta Q^{\rm T} (t)-\delta Q^*(t')\cdot Q^{\rm T} (t)\right]\right\}
  \cdot V^{\dag}(t)\cdot V(t')\right]^*\\
= -\frac{1}{8}\,\text{Tr}\int_{t_1}^{t_2} dt \int_{t_1}^{t_2} dt'\,
  E(\bar{t}, \tau)\,\left[W(\bar{t}, \tau)\,\frac{\delta \theta (t')+\delta \theta (t)}{2} + \delta Y(\bar{t}, \tau)\right]
  \cdot V^{\rm T}(t)\cdot V^*(t'),\notag
\end{multline}
where $W$ is given by \Eq{eq:WW} and $\delta Y$ is given by \Eq{eq:dY}. For $\delta\favr{\mc{B}} = \delta \mc{B}_+ + \delta \mc{B}_-$, this leads to
\begin{align}
\delta\favr{\mc{B}}
& = \frac{1}{8}\,\text{Tr} \int_{t_1}^{t_2} dt \int_{t_1}^{t_2} dt'\,
     E(\bar{t}, \tau) \left[W(\bar{t}, \tau)\,\frac{\delta \theta (t')+\delta \theta(t)}{2}+\delta Y(\bar{t}, \tau)\right] 
     \cdot \left[V^{\dag}(t)\cdot V(t')-V^{\rm T}(t)\cdot V^*(t')\right]\notag\\
& = \frac{1}{8}\,\text{Tr}\int_{t_1}^{t_2} dt \int_{t_1}^{t_2} dt'\,
    E(\bar{t}, \tau) \left[W(\bar{t}, \tau)\,\frac{\delta \theta (t')+\delta \theta(t)}{2} + \delta Y(\bar{t}, \tau)\right]
    \cdot \left[-2i \alpha(t, t')\right]\notag\\
& = - \frac{i}{4}\,\text{Tr} \int_{t_1}^{t_2} dt \int_{t_1}^{t_2} dt'\,
    E(\bar{t}, \tau) \left[W(\bar{t}, \tau)\,\frac{\delta \theta (t')+\delta \theta(t)}{2} +\delta Y(\bar{t}, \tau)\right]
    \cdot \alpha(t, t')\notag\\
& = - \frac{i}{4}\,\text{Tr}\int_{t_1}^{t_2} dt \int_{t_1}^{t_2} dt'\, e^{i \omega \left(\bar{t}\right) \tau}
    \left[W(\bar{t}, \tau)\,\delta \theta \left(\bar{t}\right)+\delta Y(\bar{t}, \tau)\right]
    \cdot \bar{\alpha} (\bar{t}, \tau) + O\left(\epsilon_\omega^2\right)\notag\\
& \approx \delta \mc{B}_{\theta}+\delta \mc{B}_Q,
\end{align}
where $\delta \mc{B}_{\theta}$ and $\delta \mc{B}_Q$ are given by \Eqs{eq:dBtheta} and \eq{eq:dBQ}, respectively. 

\end{widetext}

%%%%%%%%%%%%%%%%%%%%%%%%%%%%%%%%%%%%%%%
%\bibliography{main,foot}

\begin{thebibliography}{10}

\bibitem{foot:applic}
Practical applications of this fact include the development of advanced
  algorithms for numerical simulations that enjoy exceptional long-term
  accuracy \cite{ref:qin16, tex:stern15, ref:bridges06}.

\bibitem{book:whitham}
G.~B. Whitham, {\it Linear and Nonlinear Waves\/} (Wiley, New York, 1974).

\bibitem{book:tracy}
E.~R. Tracy, A.~J. Brizard, A.~S. Richardson, and A.~N. Kaufman, {\it Ray
  Tracing and Beyond: Phase Space Methods in Plasma Wave Theory\/} (Cambridge
  University Press, New York, 2014).

\bibitem{my:amc}
I.~Y. Dodin and N.~J. Fisch, {\it Axiomatic geometrical optics,
  Abraham-Minkowski controversy, and photon properties derived classically\/},
  Phys. Rev. A {\bf 86}, 053834 (2012).

\bibitem{my:wkin}
I.~Y. Dodin, {\it Geometric view on noneikonal waves\/}, Phys. Lett. A {\bf
  378}, 1598 (2014).

\bibitem{my:qdirac}
D.~E. Ruiz and I.~Y. Dodin, {\it Lagrangian geometrical optics of nonadiabatic
  vector waves and spin particles\/}, Phys. Lett. A {\bf 379}, 2337 (2015).

\bibitem{tex:mycovar}
D. E. Ruiz and I. Y. Dodin, {\it Extending geometrical optics: A Lagrangian theory for vector waves}, arXiv:1612.06184. 

\bibitem{my:qdiel}
D.~E. Ruiz and I.~Y. Dodin, {\it First-principles variational formulation of
  polarization effects in geometrical optics\/}, Phys. Rev. A {\bf 92}, 043805
  (2015).

\bibitem{my:qlagr}
D.~E. Ruiz and I.~Y. Dodin, {\it On the correspondence between quantum and
  classical variational principles\/}, Phys. Lett. A {\bf 379}, 2623 (2015).

\bibitem{my:qdirpond}
D.~E. Ruiz, C.~L. Ellison, and I.~Y. Dodin, {\it Relativistic ponderomotive
  Hamiltonian of a Dirac particle in a vacuum laser field\/}, Phys. Rev. A {\bf
  92}, 062124 (2015).

\bibitem{my:protation}
X.~Guan, I.~Y. Dodin, H.~Qin, J.~Liu, and N.~J. Fisch, {\it On plasma rotation
  induced by waves in tokamaks\/}, Phys. Plasmas {\bf 20}, 102105 (2013).

\bibitem{my:itervar}
I.~Y. Dodin, {\it On variational methods in the physics of plasma waves\/},
  Fusion Sci. Tech. {\bf 65}, 54 (2014).

\bibitem{my:sharm}
C.~Liu and I.~Y. Dodin, {\it Nonlinear frequency shift of electrostatic waves
  in general collisionless plasma: unifying theory of fluid and kinetic
  nonlinearities\/}, Phys. Plasmas {\bf 22}, 082117 (2015).

\bibitem{my:bgk}
I.~Y. Dodin and N.~J. Fisch, {\it Nonlinear dispersion of stationary waves in
  collisionless plasmas\/}, Phys. Rev. Lett. {\bf 107}, 035005 (2011).

\bibitem{my:trcomp}
P.~F. Schmit, I.~Y. Dodin, J.~Rocks, and N.~J. Fisch, {\it Nonlinear
  amplification and decay of phase-mixed waves in compressing plasma\/}, Phys.
  Rev. Lett. {\bf 110}, 055001 (2013).

\bibitem{tex:myqponder}
D.~E. Ruiz and I.~Y. Dodin, \textit{Ponderomotive dynamics of waves in
  quasiperiodically modulated media}, arXiv:1609.01681.

\bibitem{my:lens}
I.~Y. Dodin and N.~J. Fisch, {\it Ponderomotive forces \textit{on} waves in
  modulated media\/}, Phys. Rev. Lett. {\bf 112}, 205002 (2014).

\bibitem{my:autozen}
I.~Barth, I.~Y. Dodin, and N.~J. Fisch, {\it Ladder climbing and autoresonant
  acceleration of plasma waves\/}, Phys. Rev. Lett. {\bf 115}, 075001 (2015).

\bibitem{foot:comp}
The only exception are media with extremely simple dispersion, such as those
  described by a local refraction index. Although often presented as
  characteristic models of dielectric media, they, in fact, do not provide a
  truly representative picture of general wave dynamics.

\bibitem{foot:dense}
For a general theory, see \Refs{ref:bernstein75, ref:bornatici00,
  ref:bornatici03}. For specific examples, see, \eg \Refs{my:dense, my:mquanta,
  ref:benisti15}.
  
\bibitem{book:stix}
T.~H. Stix, {\it Waves in Plasmas\/} (AIP, New York, 1992), Chap.~8.

\bibitem{ref:benisti16}
D.~B\'enisti, {\it Envelope equation for the linear and nonlinear propagation
  of an electron plasma wave, including the effects of Landau damping,
  trapping, plasma inhomogeneity, and the change in the state of wave\/}, Phys.
  Plasmas {\bf 23}, 102105 (2016).

\bibitem{ref:dekker81}
H.~Dekker, {\it Classical and quantum mechanics of the damped harmonic
  oscillator\/}, Phys. Rep. {\bf 80}, 1 (1981).

\bibitem{ref:bateman31}
H.~Bateman, {\it On dissipative systems and related variational principles\/},
  Phys. Rev. {\bf 38}, 815 (1931).

\bibitem{ref:jimenez76}
J.~Jimenez and G.~B. Whitham, {\it An averaged Lagrangian method for
  dissipative wavetrains\/}, Proc. R. Soc. A {\bf 349}, 277 (1976).

\bibitem{ref:ibragimov06}
N.~H. Ibragimov, {\it Integrating factors, adjoint equations and
  Lagrangians\/}, J. Math. Anal. Appl. {\bf 318}, 742 (2006).

\bibitem{ref:galley13}
C.~R. Galley, {\it Classical mechanics of nonconservative systems\/}, Phys.
  Rev. Lett. {\bf 110}, 174301 (2013).

\bibitem{ref:caldirola41}
P.~Caldirola, {\it Forze non conservative nella meccanica quantistica\/}, Nuovo
  Cimento {\bf 18}, 393 (1941).

\bibitem{ref:kanai48}
E.~Kanai, {\it On the quantization of the dissipative systems\/}, Prog. Theor.
  Phys. {\bf 3}, 440 (1948).

\bibitem{ref:herrera86}
L.~Herrera, L.~N\'u$\tilde{\text{n}}$ez, A.~Pati$\tilde{\text{n}}$o, and
  H.~Rago, {\it A variational principle and the classical and quantum mechanics
  of the damped harmonic oscillator\/}, Am. J. Phys. {\bf 54}, 273 (1986).

\bibitem{ref:chandrasekar07}
V.~K. Chandrasekar, M.~Senthilvelan, and M.~Lakshmanan, {\it On the Lagrangian
  and Hamiltonian description of the damped linear harmonic oscillator\/}, J.
  Math. Phys. {\bf 48}, 032701 (2007).

\bibitem{ref:bender07}
C.~M. Bender, M.~Gianfreda, N.~Hassanpour, and H.~F. Jones, {\it Comment on
  ``On the Lagrangian and Hamiltonian description of the damped linear harmonic
  oscillator'' [J. Math. Phys. 48, 032701 (2007)]\/}, J. Math. Phys. {\bf 57},
  084101 (2016).

\bibitem{ref:musielak08}
Z.~E. Musielak, {\it Standard and non-standard Lagrangians for dissipative
  dynamical systems with variable coefficients\/}, J. Phys. A: Math. Theor.
  {\bf 41}, 055205 (2008).

\bibitem{ref:lakshmanan13}
M.~Lakshmanan and V.~K. Chandrasekar, {\it Generating finite dimensional
  integrable nonlinear dynamical systems\/}, Eur. Phys. J. Special Topics {\bf
  222}, 665 (2013).

\bibitem{ref:riewe97}
F.~Riewe, {\it Mechanics with fractional derivatives\/}, Phys. Rev. E {\bf 55},
  3581 (1997).

\bibitem{ref:rabei04}
E.~M. Rabei, T.~S. Alhalholy, and A.~A. Taani, {\it On Hamiltonian formulation
  of non-conservative systems\/}, Turk. J. Phys. {\bf 28}, 213 (2004).

\bibitem{ref:frederico16}
G.~S.~F. Frederico and M.~J. Lazo, {\it Fractional Noether's theorem with
  classical and Caputo derivatives: constants of motion for non-conservative
  systems\/}, Nonlinear Dyn. {\bf 85}, 839 (2016).

\bibitem{ref:bruneau02}
L.~Bruneau and S.~De Bi\'evre, {\it A Hamiltonian model for linear friction in
  a homogeneous medium\/}, Commun. Math. Phys. {\bf 229}, 511 (2002).

\bibitem{ref:montgomery14}
S.~Montgomery-Smith, {\it Hamiltonians representing equations of motion with
  damping due to friction\/}, Electron. J. Differential Equations {\bf 2014}, 1
  (2014).

\bibitem{ref:virga15}
E.~G. Virga, {\it Rayleigh-Lagrange formalism for classical dissipative
  systems\/}, Phys. Rev. E {\bf 91}, 013203 (2015).

\bibitem{arX:bravetti16}
A.~Bravetti, H.~Cruz, and D.~Tapias, {\it Contact Hamiltonian mechanics\/},
  arXiv:1604.08266.

\bibitem{ref:huttner92}
B.~Huttner and S.~M. Barnett, {\it Quantization of the electromagnetic field in
  dielectrics\/}, Phys. Rev. {\bf 46}, 4306 (1992).

\bibitem{ref:celeghini92}
E.~Celeghini, M.~Rasetti, and G.~Vitiello, {\it Quantum dissipation\/}, Ann.
  Phys. {\bf 215}, 156 (1992).

\bibitem{ref:gruner95}
T.~Gruner and D.-G. Welsch, {\it Correlation of radiation-field ground-state
  fluctuations in a dispersive and lossy dielectric\/}, Phys. Rev. A {\bf 51},
  3246 (1995).

\bibitem{ref:bolivar98}
A.~O. Bolivar, {\it Quantization of non-Hamiltonian physical systems\/}, Phys.
  Rev. A {\bf 58}, 4330 (1998).

\bibitem{ref:bechler99}
A.~Bechler, {\it Quantum electrodynamics of the dispersive dielectric medium --
  a path integral approach\/}, J. Mod. Opt. {\bf 46}, 901 (1999).

\bibitem{ref:wonderen04}
A.~J. van Wonderen and L.~G. Suttorp, {\it The oscillator model for dissipative
  QED in an inhomogeneous dielectric\/}, J. Phys. A: Math. Gen. {\bf 37}, 11101
  (2004).

\bibitem{ref:bechler06}
A.~Bechler, {\it Path-integral quantization of the electromagnetic field in the
  Hopfield dielectric beyond dipole approximation\/}, J. Phys. A: Math. Gen.
  {\bf 39}, 13553 (2006).

\bibitem{ref:suttorp07}
L.~G. Suttorp, {\it Field quantization in inhomogeneous anisotropic dielectrics
  with spatio-temporal dispersion\/}, J. Phys. A: Math. Theor. {\bf 40}, 3697
  (2007).

\bibitem{ref:philbin10}
T.~G. Philbin, {\it Canonical quantization of macroscopic electromagnetism\/},
  New J. Phys. {\bf 12}, 123008 (2010).

\bibitem{ref:behunin11}
R.~O. Behunin and B.-L. Hu, {\it Nonequilibrium forces between atoms and
  dielectrics mediated by a quantum field\/}, Phys. Rev. A {\bf 84}, 012902
  (2011).

\bibitem{ref:horsley11}
S.~A.~R. Horsley, {\it Consistency of certain constitutive relations with
  quantum electromagnetism\/}, Phys. Rev. A {\bf 84}, 063822 (2011).

\bibitem{ref:braun11}
M.~A. Braun, {\it QED in dispersive and absorptive media\/}, Theor. Math. Phys.
  {\bf 169}, 1413 (2011).

\bibitem{arX:churchill15}
R.~J. Churchill and T.~G. Philbin, {\it Absorption in dielectric models\/},
  arXiv:1508.04666.

\bibitem{foot:dewar}
R. L. Dewar, R. F. Abdullatif, and G. G. Sangeetha, \textit{Complex nonlinear
  Lagrangian for the Hasegawa-Mima equation}, in Proceedings of the 20th IAEA
  Fusion Energy Conference (2004), TH/P6-1,
  \url{http://www-naweb.iaea.org/napc/physics/fec/fec2004/datasets/TH_P6-1.html}.

\bibitem{foot:ext}
In a nutshell, for any equation $F[q] = 0$, the function $rF[q]$ can serve as a
  Lagrangian if both $q$ and $r$ are treated as independent variables.

\bibitem{ref:mcdonald85}
S.~W. McDonald and A.~N. Kaufman, {\it Weyl representation for electromagnetic
  waves: The wave kinetic equation\/}, Phys. Rev. A {\bf 32}, 1708 (1985).

\bibitem{book:mendonca}
J.~T. Mendon\c{c}a, {\it Theory of Photon Acceleration\/} (IOP, Philadelphia,
  2000).

\bibitem{tex:myzonal}
D. E. Ruiz, J. B. Parker, E. L. Shi, and I. Y. Dodin, \textit{Zonal-flow dynamics from a phase-space perspective}, Phys. Plasmas {\bf 23}, 122304 (2016).  

\bibitem{ref:bornatici00}
M.~Bornatici and Yu.~A. Kravtsov, {\it Comparative analysis of two formulations
  of geometrical optics. The effective dielectric tensor\/}, Plasma Phys.
  Control. Fusion {\bf 42}, 255 (2000).

\bibitem{ref:bornatici03}
M.~Bornatici and O.~Maj, {\it Geometrical optics response tensors and the
  transport of the wave energy density\/}, Plasma Phys. Control. Fusion {\bf
  45}, 1511 (2003).

\bibitem{ref:balakin15}
A.~A. Balakin and E.~D. Gospodchikov, {\it Operator formalism for permittivity
  tensor in smoothly inhomogeneous media with spatial dispersion\/}, J. Phys.
  B: At. Mol. Opt. Phys. {\bf 48}, 215701 (2015).

\bibitem{book:landau1}
L.~D. Landau and E.~M. Lifshitz, {\it Mechanics\/} (Butterworth-Heinemann,
  Oxford, 1976).

\bibitem{book:lanczos}
C.~Lanczos, {\it The Variational Principles of Mechanics\/} (Univ. Toronto
  Press, Toronto, 1964), second

\bibitem{foot:mouchet}
A related discussion can be found in \Ref{arX:mouchet15}.

\bibitem{foot:neg}
An example of a negative-energy mode is a mode governed by a Lagrangian
  $\smash{-\dot{\xi}^2/2 + \Omega^2\xi^2/2}$, which is minus the Lagrangian of
  a usual, positive-energy mode.

\bibitem{foot:contrast}
This implies $M$ \textit{real} BC per each end point, as in the original
  problem (\Sec{sec:coupled}). This is in contrast with \Ref{my:wkin}, where
  $M$ \textit{complex} BC were assumed per end point in the quantumlike
  representation [namely, $\delta\psi(t_{1,2}) = 0$], so the problem was
  overdefined.

\bibitem{ref:domingos84}
J.~M. Domingos and M.~H. Caldeira, {\it Self-adjointness of momentum operators
  in generalized coordinates\/}, Found. Phys. {\bf 14}, 147 (1984).

\bibitem{foot:weyl}
A brief overview of the Weyl calculus can be found, for instance, in
  \Refs{tex:myzonal, ref:imre67}. Here, we only notice that, for an operator
  $A$ given by $A \circ f = \smash{\int^{\infty}_{-\infty}} A(t, t')
  f(t')\,dt'$, its Weyl image is $\smash{A_W(\bar{t}, \omega) \doteq
  \int^{\infty}_{-\infty} A(\bar{t} + \tau/2, \bar{t} - \tau/2)\,e^{i \omega
  \tau}\,d\tau}$. The polarizability operator can be brought to the form
  assumed for $A \circ f$ by making two replacements in \Eq{eq:alphaq}: (i)
  $t_1 \to -\infty$, which is justified provided we consider the operator on
  functions that are identically zero at $t < t_1$; and (ii) $t_2 \to +\infty$,
  which is justified by the presence of the Heaviside function in the
  integrand.

\bibitem{foot:adiab}
As asymptotic approximation of $\smash{\mcu{J}[f] \doteq \int_a^b
  g(t)\,e^{i\vartheta(t)} \,dt}$, where $g$ evolves at some rate
  $\smash{\mc{T}^{-1} \ll \omega \doteq - \dot{\vartheta}}$, can be found via
  integration by parts. This gives $\mcu{J}[g] = \smash{\int_a^b} [(d/dt)(ig
  e^{i\vartheta}/\omega) - i g'(t)/\omega]\,dt$ or, equivalently, $\mcu{J}[g] =
  \smash{(ig e^{i\vartheta}/\omega)|_a^b} - \smash{\mcu{J}[i g'(t)/\omega]}$,
  where the functional on the right is smaller than the functional on the left
  by a factor $O(\omega \mc{T}) \gg 1$. The calculation can also be iterated.
  Then, $\mcu{J}[g]$ can be represented as an asymptotic series of terms that
  are determined by $f$, $\theta$ and their derivatives evaluated locally at
  the ends of the integration domain.

\bibitem{my:kchi}
I.~Y. Dodin and N.~J. Fisch, {\it On generalizing the $K$-$\chi$ theorem\/},
  Phys. Lett. A {\bf 374}, 3472 (2010).

\bibitem{ref:kaufman87}
A.~N. Kaufman, {\it Phase-space-Lagrangian action principle and the generalized
  $K$-$\chi$ theorem\/}, Phys. Rev. A {\bf 36}, 982 (1987).

\bibitem{ref:cary77}
J.~R. Cary and A.~N. Kaufman, {\it Ponderomotive force and linear
  susceptibility in Vlasov plasma\/}, Phys. Rev. Lett. {\bf 39}, 402 (1977).

\bibitem{foot:act}
In linear systems, $I = \mc{E}/\omega$, where $\mc{E}$ is the energy. Thus, the
  action conservation law that we presented is also known as the conservation
  of an adiabatic invariant $\mc{E}/\omega$. For details, see, \eg
  \Ref{my:amc}.
  
\bibitem{book:kravtsov}
Yu.~A. Kravtsov and Yu.~I. Orlov, {\it Geometrical optics of inhomogeneous
  media\/} (Springer-Verlag, New York, 1990).

\bibitem{my:acti}
I.~Y. Dodin and N.~J. Fisch, {\it Adiabatic nonlinear waves with trapped
  particles: I. General formalism\/}, Phys. Plasmas {\bf 19}, 012102 (2012).

\bibitem{foot:appl}
For applications of \Eq{eq:aeffl}, see, \eg the analytical models of nonlinear
  plasma waves developed in \Refs{my:itervar, my:sharm, my:bgk, my:trcomp}.

\bibitem{my:mneg}
I.~Y. Dodin and N.~J. Fisch, {\it Positive and negative effective mass of
  classical particles in oscillatory and static fields\/}, Phys. Rev. E {\bf
  77}, 036402 (2008).

\bibitem{ref:qin16}
H.~Qin, J.Liu, J.~Xiao, R.~Zhang, Y.~He, Y.~Wang, Y.~Sun, J.~W. Burby,
  L~Ellison, and Y~Zhou, {\it Canonical symplectic particle-in-cell method for
  long-term large-scale simulations of the Vlasov-Maxwell equations\/}, Nucl.
  Fusion {\bf 56}, 014001 (2015).

\bibitem{tex:stern15}
A. Stern, Y. Tong, M. Desbrun, and J.~E. Marsden, \textit{Geometric
  computational electrodynamics with variational integrators and discrete
  differential forms}, in \textit{Geometry, Mechanics, and Dynamics: The Legacy
  of Jerry Marsden}, edited by D. E. Chang, D. D. Holm, G. Patrick, and T.
  Ratiu (Springer, New York, 2015), p. 437.

\bibitem{ref:bridges06}
T.~J. Bridges and S.~Reich, {\it Numerical methods for Hamiltonian PDEs\/}, J.
  Phys. A: Math. Gen. {\bf 39}, 5287 (2006).

\bibitem{ref:bernstein75}
I.~B. Bernstein, {\it Geometric optics in space and time varying plasmas. I\/},
  Phys. Fluids {\bf 18}, 320 (1975).

\bibitem{my:dense}
I.~Y. Dodin, V.~I. Geyko, and N.~J. Fisch, {\it Langmuir wave linear evolution
  in inhomogeneous nonstationary anisotropic plasma\/}, Phys. Plasmas {\bf 16},
  112101 (2009).

\bibitem{my:mquanta}
I.~Y. Dodin and N.~J. Fisch, {\it On the evolution of linear waves in
  cosmological plasmas\/}, Phys. Rev. D {\bf 82}, 044044 (2010).

\bibitem{ref:benisti15}
D.~B\'enisti, {\it Kinetic description of linear wave propagation in
  inhomogeneous, nonstationary, anisotropic, weakly magnetized, and collisional
  plasma\/}, Phys. Plasmas {\bf 22}, 072106 (2015).

\bibitem{arX:mouchet15}
A.~Mouchet, {\it Applications of Noether conservation theorem to Hamiltonian
  systems\/}, arXiv:1601.01610.

\bibitem{ref:imre67}
K.~Imre, E.~\"Ozizmir, M.~Rosenbaum, and P.~F. Zweifel, {\it Wigner method in
  quantum statistical mechanics\/}, J. Math. Phys. {\bf 8}, 1097 (1967).

\end{thebibliography}

\end{document}